\documentclass[journal]{IEEEtran}

\ifCLASSINFOpdf
\else
\fi
\usepackage[cmex10]{amsmath}
\usepackage{amsfonts}
\usepackage{amssymb}
\usepackage{empheq}

\usepackage{url}

\usepackage{booktabs}

\begin{document}
%
\title{Modeling and Performance of Contact-Free Discharge Systems for Space Inertial Sensors}
\author{Tobias~Ziegler,
        Patrick~Bergner, 
				Gerald~Hechenblaikner,
				Nico~Brandt,
        and~Walter~Fichter
\thanks{T.~Ziegler, P.~Bergner, G.~Hechenblaikner, and N.~Brandt are with Astrium GmbH, Science Missions Department (AED41), 88039 Friedrichshafen, Germany e-mail: tobias.ziegler@astrium.eads.net.}
\thanks{W. Fichter is managing director of the Institute of Flight Mechanics and Control, University of Stuttgart, Pfaffenwaldring 7a, 70569 Stuttgart, Germany.}
}

\maketitle

\begin{abstract}
This article presents a detailed overview and assessment of contact-free UV light discharge systems (UVDS) needed to control the variable electric charge level of free-flying test masses which are part of high precision inertial sensors in space. A comprehensive numerical analysis approach on the basis of experimental data is detailed. This includes UV light ray tracing, the computation of time variant electric fields inside the complex inertial sensor geometry, and the simulation of individual photo-electron trajectories. 
Subsequent data analysis allows to determine key parameters to set up an analytical discharge model. Such a model is an essential system engineering tool needed for requirement breakdown and subsystem specification, performance budgeting, on-board charge control software development, and instrument modeling within spacecraft end-to-end performance simulators. 
Different types of UVDS design concepts are presented and assessed regarding their robustness and performance. Critical hardware aspects like electron emission from air-contaminated surfaces, interfaces with other subsystems, and spacecraft operations are considered. The focus is on the modeling and performance evaluation of the existing UVDS on board LISA Pathfinder, an ESA technology demonstrator spacecraft to be launched in 2014. The results have motivated the design of a more robust discharge system concept for cubical test mass inertial sensors for future space missions. The developed analysis tools have been used for design optimization and performance assessment of the proposed design. A significant improvement of relevant robustness and performance figures has been achieved. 
\end{abstract}


\begin{IEEEkeywords}
drag-free control, charge management, discharge, inertial sensor, ray tracing, electron tracing, photoemission, UV light, LISA, NGO, LISA Pathfinder.
\end{IEEEkeywords}

\IEEEpeerreviewmaketitle

\section*{Nomenclature}
\noindent\begin{tabular}{@{}lcl@{}}
AC  								  &=& Alternating current\\
AIT  								  &=& Assembly, integration, and test\\
BRDF 								  &=& Bidirectional reflectance distribution function\\
CAD										&=& Computer-aided design\\
DC  								  &=& Direct current\\
DFACS                 &=& Drag-free and attitude control system\\
DLR										&=& German aerospace center\\
DoF  								  &=& Degree of freedom\\
EH  								  &=& Electrode housing with electrodes\\
ESA  								  &=& European Space Agency\\
GCR  								  &=& Galactic cosmic ray\\
LED                   &=& Light-emitting diode\\
LISA  								&=& Laser Interferometer Space Antenna\\
LPF  									&=& LISA Pathfinder\\
LTP  									&=& LISA Technology Package\\
MBW  									&=& Measurement bandwidth\\
NGO                   &=& New Gravitational Wave Observatory\\
PDF  									&=& Probability density function\\
PTB										&=& National Metrology Institute of Germany\\
SEP   								&=& Solar energetic particle\\
SC 								 	  &=& Spacecraft\\
TM   									&=& Test mass\\
UV										&=& Ultra-violet\\
UVDS									&=& UV light discharge system\\
UVLF									&=& UV light feedthrough
\end{tabular}
\\

\section{Introduction}
\label{sec:introduction}
%
%
%
%

\subsection{Problem Formulation}
\label{subsec:problem_formulation}
\IEEEPARstart{I}{nertial} 
sensors are key components of scientific space missions aimed at measuring the effects of spacetime curvature caused by celestial bodies. Spacetime curvature can be measured directly using at least two free-falling\footnotemark\footnotetext{Free-falling particles are not disturbed by any forces and move along their geodesic lines in curved space-time \cite{misner1973gravitation}.} test masses as gravity references in space, by observing the distance variation between these reference points. The variation between a test mass and its hosting spacecraft is used as reference for the spacecraft drag-free control system \cite{lange_1964dragfreesatellite}\cite{fichter_LPF_DFACS2} needed to suppress the spurious non-gravitational forces acting on the shielding spacecraft. 

The envisaged space-born gravitational wave detector LISA/NGO\footnotemark\footnotetext{Since 2011, in the framework of an ESA only mission, LISA is called NGO.} \cite{bell_LISAandLPFOverview}\cite{esa_yb2012} relies on the direct measurement of spacetime curvature, using laser interferometry for the distance measurement and high precision inertial sensors with cubical test masses as the gravity references. A description of a modern high precision inertial sensor for space is given in Appendix~\ref{subsec:ISdescription}. The currently assembled LISA Pathfinder (LPF) spacecraft is an ESA/NASA technology demonstrator (to be launched in 2014 and operated in an orbit around the L1 Lagrange point), aimed at testing novel LISA technologies \cite{racca_TheLPFMisson}. Its scientific payload, the LISA Technology Package \cite{gerndt_LTPoverview} is, among other instruments, equipped with two high precision inertial sensors.
The sensors must demonstrate a level of residual, non-gravitational acceleration noise in the order of $10^{-14}$\,m/s$^2$/Hz$^{1/2}$ or below, down to millihertz frequencies\footnotemark\footnotetext{The LISA Pathfinder requirement on differential acceleration noise between two test masses along the sensitive axis is $3\cdot10^{-14}$\,m/s$^2$/Hz$^{1/2}$ in the measurement bandwidth (MBW) from $1-30$\,mHz. The LISA acceleration noise requirement is $3\cdot10^{-15}$\,m/s$^2$/Hz$^{1/2}$ in the MBW from $0.1-100$\,mHz.}. 

The cubical test masses of the LPF and the LISA/NGO inertial sensors have no mechanical contact with their surrounding housing structures. Ideally, in the absence of any non-gravitational forces, the test masses should move along geodesic lines. However, in the real instruments, pure geodesic test mass motion is impaired by non-gravitational forces \cite{antonucci_performance} that accelerate the test masses away from their geodesic lines.

A major source of non-gravitational disturbance noise is caused by the net electric charge $\text{Q}_\text{TM}$ accumulating on the test masses. Sources of test mass charging are:
\begin{enumerate}
\item Contact electrification \cite{contact_electrification} due to test mass release into free-flight from the release mechanism (see Appendix~\ref{subsec:ISdescription}).
\item Energetic, charged particles such as galactic cosmic rays (GCRs) and large amounts of high-energy plasma, periodically expelled by the sun in solar energetic particle (SEP) events. These particles penetrate the spacecraft shielding and eject or deposit charges on the test mass \cite{vocca_LISAChargingSimulation}\cite{araujo_LISAChargingSimulation}.
\end{enumerate}

Taking LISA Pathfinder as an example, from the first effect, it is expected that a charge level up to $\pm2.35\cdot10^8$ elementary charges ($\text{e}$) with arbitrary sign will deposit on a test mass \cite{vitale_LPFcontactpotential}. This corresponds to a test mass potential $\text{V}_\text{TM}=\pm1.1$\,V relative to the grounded housing surfaces.
From the second type of effects, a charge rate of about $+50$\,e/s is expected due to GCRs for solar minimum conditions which corresponds to approximately $+4.3\cdot10^6\,\text{e}$ ($+20$\,mV) within 24\,h. Furthermore, about $+5\cdot10^6$ elementary charges accumulate due to one typical, \emph{small} SEP event \cite{wass_LPFChargingSimulation}. The probability of such an event occurring in a 24\,h period is estimated to be $2.4$\,\% (calculated from the data reported in \cite[Fig.\,3]{wass_LPFChargingSimulation} for the year $2011$). 

Thus, $24$\,h after test mass release, up to $+2.39\cdot10^8\,\text{e}$ ($\text{V}_\text{TM}=+1.12$\,V) may deposit on the test mass---assuming a \emph{positive} charge level has been accumulated on the test mass due to contact electrification after release. With the same argument, after six month without re-grabbing the test mass, a potential of up to $+4.9$\,V might be present on the test mass (assuming $4$ smaller SEP events).\\  

\begin{figure}[t]
\begin{center}
\includegraphics[width=0.49\textwidth]{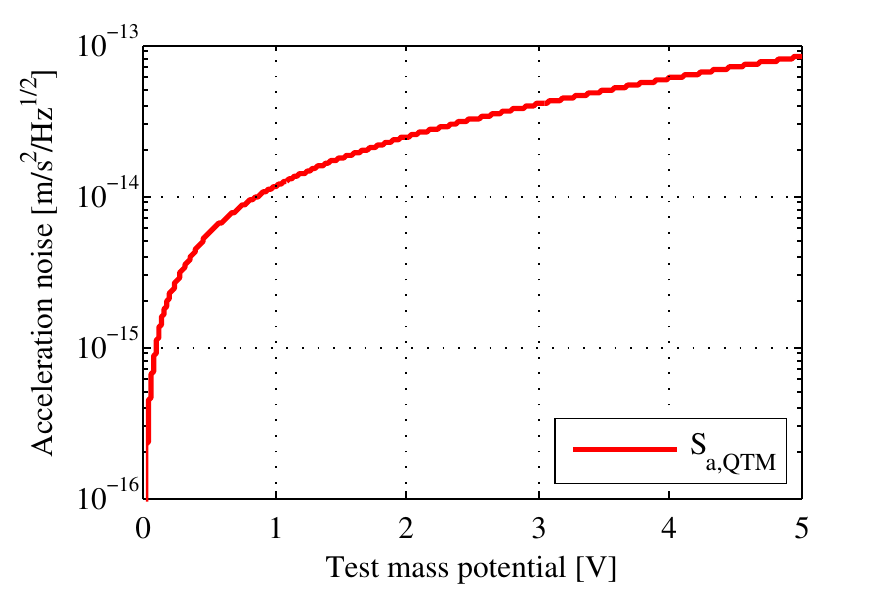} 
\end{center}
\caption{\label{fig:chargeeffects} Linear spectral density of acceleration noise effects due to coupling with accumulated test mass charge at $1$\,mHz. The budget is shown for the reference inertial sensor described in Appendix~\ref{subsec:ISdescription}.}
\end{figure}

Figure~\ref{fig:chargeeffects} shows the contribution of all acceleration noise effects on the test mass at 1\,mHz, caused by a coupling with the test mass charge $\text{Q}_\text{TM}$. The linear spectral density $\text{S}_\text{a,QTM}$ of the acceleration noise at $1$\,mHz is plotted versus increasing test mass potential $\text{V}_\text{TM}=\text{Q}_\text{TM}/\text{C}_\text{tot}$, where $\text{C}_\text{tot}$ is the total electrostatic capacitance between the test mass and its surrounding housing. $\text{S}_\text{a,QTM}$ is the root-sum-square value of multiple contributors that couple with $\text{Q}_\text{TM}$ (e.g., coupling of voltage noise from electrostatic actuation, stray voltage fluctuations, increase of stiffness coupling with the control jitter, etc.)\footnotemark\footnotetext{Additional acceleration noise effects at $1$\,mHz due to charged particle impacts from the environment and charge control actuation are not considered in $\text{S}_\text{a,QTM}$ since it contains only effects that couple with \emph{accumulated} DC test mass charge $\text{Q}_\text{TM}$.}.
Note that the linear spectral density of the \emph{total} residual test mass acceleration noise is given by the sum of the charge related effects and various other contributors \cite{antonucci_performance}\cite{antonucci_2012interaction}. However, Figure~\ref{fig:chargeeffects} shows that with increasing test mass charge, also $\text{S}_\text{a,QTM}$ increases, such that, at $\text{V}_\text{TM}=900$\,mV, the acceleration noise at $1$\,mHz is already at $10^{-14}$\,m/s$^2$/Hz$^{1/2}$---only due to test mass charge related effects. 

Since the LISA Pathfinder mission goal is to properly characterize all effects causing acceleration noise below $10^{-14}$\,m/s$^2$/Hz$^{1/2}$, the test mass charge has to be much smaller to limit its acceleration noise contribution. Moreover, the LISA/NGO requirement on the residual test mass potential is $0.5$\,mV ($\text{Q}_\text{TM}=1\cdot10^5$\,e) in order not to disturb the science measurements. The verification of the required LISA/NGO charge level control accuracy is one of the experiments to be demonstrated with LISA Pathfinder \cite{esa_lpf_emp}.

Thus, in order to meet the stringent acceleration noise requirements, positive and negative charges must be removed from the inertial sensor test masses (``bipolar discharging''). Charge control by using a thin conducting gold wire as an electrical connection between the test mass and the electrode housing structure (e.g., applied in modern accelerometers as used in GOCE \cite{marque_GOCE} or MICROSCOPE \cite{hudson_microscope_preflight_performance}) is not an option for inertial sensors with residual acceleration noise requirements in the order of $10^{-14}$\,m/s$^2$/Hz$^{1/2}$. According to \cite{willemenot_goldwiredamping}, such wires introduce increased stiffness and damping. The damping force itself is negligible, but the thermal noise it causes (fluctuation-dissipation theorem) limits the performance of an LPF and LISA/NGO like inertial sensor to $4.5\cdot10^{-14}$\,m/s$^2$/Hz$^{1/2}$ at $1$\,mHz (i.e., already above the total noise requirement).


As a consequence, the test mass charge has to be controlled without any mechanical contact to the surrounding housing structure. This can be achieved by using ultra-violet (UV) light via the photoelectric effect\footnotemark\footnotetext{Also alternative techniques for the implementation of contact-free test mass discharge systems have been discussed. These techniques make use of radioactive sources, ion sources, or field emission cathodes (see e.g. \cite{buchman_charge_control_fieldemission_cathodes}).}.

\subsection{Previous work and significance this article}
\label{subsec:previouswork} 
A contact-free, photoemission based UV discharge system has been used to control the charge of the four electrostatically suspended spherical test masses on-board Gravity Probe B \cite{buchman_GPB_charge_measurement_control}. The surfaces of the test mass and of a dedicated charge control electrode (integrated in the surrounding housing structure) have been illuminated with UV light; the direction of the induced photo-currents from both surfaces has been controlled by application of either positive or negative DC bias voltages to the charge control electrode (3\,V with respect to the test mass surface). 

An early design of the discharge system for the cubical, gold coated test masses of the LISA Pathfinder and LISA inertial sensors is described in \cite{sumner_description_LISA_chargingdischarging}. 
A modification of the UV light injection angle into the sensor from the original design is described in \cite{shaul_charge_management_LISA_and_LPF}. An analysis of the light injection modification and the influence on achievable discharge performance and robustness is not reported therein. 

In \cite{sumner_LISA_and_LPF_charging} a simulation of closed-loop drag-free and attitude control including charge control operations is mentioned; the description of the used discharge system model is omitted. In \cite{hollington_diss} a simplified discharge model is reported, similar to the model previously developed by the authors \cite{ziegler_TN_dischargemodel}. The simplified model treats specific pairs of housing surfaces and their adjacent surfaces on the test mass as parallel plate capacitors, where a uniform electric field (neglecting edge effects) is assumed. Photo-electron emission is assumed perpendicular to the emitting surface and therefore constrains the released electrons to the region between the two adjacent surfaces. However, a major part of the electron emission within the LPF inertial sensor happens at corner regions, where the illuminated housing parts have no adjacent test mass surfaces and the electrical fields can not be described by parallel plate capacitors. 
 
Discharge concepts using UV LEDs are described in various publications (e.g., \cite{pollack_chargemanagement_usingLED}\cite{kexun_LEDandACdischarging}). However, a design concept for the cubical, gold coated test-masses of the LPF and LISA/NGO inertial sensors as well as a model-based design justification including robustness and performance assessments (e.g. considering the effects of photoemission from air-contaminated surfaces, UV light ray tracing in complex sensor geometries, and calculation of electron transition ratios between relevant surfaces) is not presented.\\


This article summarizes the operational principle of UV light discharge systems and gives an overview of different design concepts in Section~\ref{sec:dischargesystemdescription}. A comprehensive toolbox for the mathematical modeling of such discharge systems is presented in Section~\ref{discharge_model}. It considers crucial aspects like photoemission from air-contaminated surfaces with realistic models for the kinetic energy distribution and the angular distribution of the photo-electrons, the implementation of detailed geometrical sensor features, electron propagation through complex and time-variant electric fields that have been computed with a finite-element tool, as well as UV light scattering measurements from rough surfaces. The toolbox is used to derive an analytical model, needed for the breakdown of discharge rate level requirements down to requirements on subsystem level. Such a model is also essential for the development of the on-board charge control software, for the simulation of closed-loop charge control performance by means of spacecraft end-to-end simulators (see \cite{ziegler_PrinciplesLPFDischargeSystem}), and for performance budgeting and control during the spacecraft integration and verification process. In Section~\ref{sec:desinganalysis}, the developed tools are used to analyze the LPF discharge system design \cite{sumner_description_LISA_chargingdischarging}\cite{shaul_charge_management_LISA_and_LPF} and to justify important design modifications.

Aided by the toolbox, the robustness of different design concepts can be conveniently evaluated. A more robust, model-based design for future space inertial sensors is presented in Section~\ref{sec:LISAdesign}, and also evaluated in terms of robustness and performance. The design concept makes use of synchronized, high-frequency switching of UV LEDs, combined with an ideal light injection into the inertial sensor. In this way, the already existing high-frequency voltages (applied by the injection electrodes for electrostatic measurement purposes) will ideally assist robust and bipolar test mass discharging.

\section{Overview of UV Light Discharge Systems}
\label{sec:dischargesystemdescription}
This sections describes the basic principles of contact-free UV light discharge systems for inertial sensors and introduces different design concepts. Critical aspects like the sensitivity of the photoemission process on air-contaminated sensor surfaces \cite{hechenblaikner_photoemission} and the UV wavelength are addressed.
		
\subsection{Principle of Contact-Free UV Light Discharge Systems}
\label{subsec:UVdischargesystems}
\begin{figure}[!t]
\centering
\includegraphics[width=0.46\textwidth]{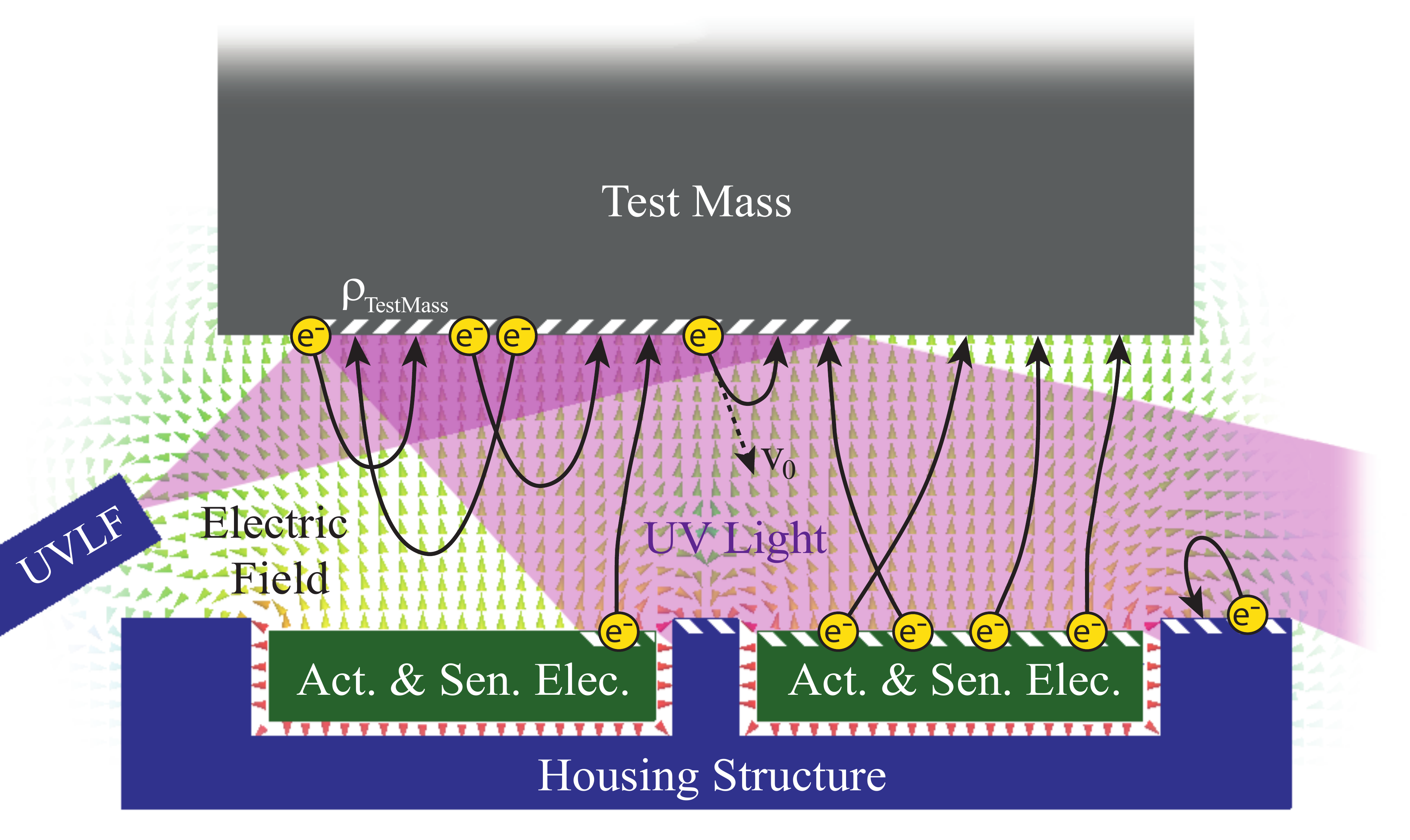}
\caption{Principle of test mass discharging shown for a negative discharge rate.
The key parameters are the absorbed UV light on EH and TM, the amount of electrons released per absorbed UV light (quantum yield), and the photo-electron transition ratios between EH and TM surfaces. The transition ratios are strongly influenced by the electric fields and the initial velocity vector $v_0$ of the photo-electrons.}
\label{fig:principle_discharging}
\end{figure}

The principle to discharge a free-flying test mass using UV light is illustrated in Figure~\ref{fig:principle_discharging}. The basic idea is to illuminate specific surfaces of the test mass and the surrounding housing with UV light such that photo-electrons are released from these surfaces via the photoelectric effect. A positive test mass discharge rate $\dot{Q}_\text{TM}>0$ is obtained when the electron flow from the test mass to the housing dominates the electron flow from the housing to the test mass. A negative discharge rate $\dot{Q}_\text{TM}<0$ is obtained when the electron flow from the housing to the test mass is dominating.
The UV light is injected via UV light feedthroughs (UVLF). Usually, a significant amount of light is reflected between individual inertial sensor surfaces. Thus, in addition to the primary illuminated surfaces, also other surfaces absorb UV light. The fraction of injected light absorbed by an arbitrary surface $i$ (taking multiple reflections into account) is denoted $\rho_i$. In Figure~\ref{fig:principle_discharging}, the totally absorbed light on the test mass is indicated by the shaded area denoted $\rho_\text{TestMass}$. Only UV light that is absorbed by a surface contributes to the photoelectric effect.

In principle, when the energy of the absorbed light equals the material work function, photoemission occurs for those electrons occupying the highest energy level (Fermi). Increasing the photon energy even further will also lead to emission of valence electrons below the Fermi level, which then have sufficient energy to be excited above the vacuum level \cite{cahen_electron_energetics_at_surfaces}. According to Fowler's derivation \cite{fowler_AnalysisOfPhotoelectricSensitivityCurves}, the photo-current $I$ for zero Kelvin should scale around the emission threshold (i.e., when electrons are removed from energy levels close to the Fermi energy) as
\begin{equation}
\label{eg:fowler}
I\,=\,A\cdot\left(h\,\nu-\Phi-eV_{i}\right)^2,
\end{equation}
for $(h\nu-\Phi)>0$ and $(h\nu-\Phi-eV_{i})>0$. In Eq.~\ref{eg:fowler}, $h\nu$ is the photon energy, $A$ is a constant, $\Phi$ the material work function, and $V_\text{i}$ an arbitrary voltage that is applied between the emitting and the receiving surface. Hence, the photo-current from one surface to another constitutes the total of all emitted electrons up to a certain energy of the voltage $V_\text{i}$. The electric field caused by the voltage $V_\text{i}$ enhances or attenuates the electron flow. $V_\text{i}$ originates from AC and DC voltages applied to individual electrodes (see eqs.~\ref{eqn:suspension_voltages}--\ref{eq:injectionvoltage} in Appendix~\ref{subsec:ISdescription}), and from the test mass potential V$_\text{TM}$. The distribution of kinetic energies normal to the material surface is approximately\footnotemark\footnotetext{Equation~\ref{eg:fowler} and Fowler's law apply to the normal energy only; the total energy distribution is given by DuBridge's equation, according to \cite{dubridge_theory_energy_distribution_photoelectrons}.} given by the derivative of Eq.~\ref{eg:fowler}. The exact kinetic energy distribution of the photo-electrons is, however, obtained from the analysis of dedicated measurement campaigns \cite{hechenblaikner_photoemission}. The photo-current between representative sample surfaces and a collecting anode has been measured while, a bias voltages was stepwise increased until the photo-current was entirely suppressed. In this way, the \emph{total} energy distribution can be derived; it includes important features like the distribution at about $300$\,K (which is within the operating temperature range of the inertial sensor) and the low-energy modification due to energy dependent transmission barriers. A typical kinetic energy distribution (derived for gold surfaces under $253.6$\,nm UV illumination) is shown in Section~\ref{discharge_model_electrontracer}. 

Whether individual surfaces contribute to positive or negative test mass discharge rates depends on the trajectories of the photo-electrons in the volume between the test mass and the housing structure. In addition to the present electric fields, the trajectories depend on the position, the direction, and the kinetic energy of the photo-electrons. In Figure~\ref{fig:principle_discharging}, the initial state of the photo-electrons is indicated by the velocity vector v$_\text{0}$.

The number of released electrons from a surface per absorbed light is characterized by the \emph{quantum yield}. The total quantum yield comprises contributions from the bulk material on the one hand and effects relating to the material surface on the other hand. 
The photo-current of Eq.~\ref{eg:fowler} only takes bulk emission into account. Phenomena of the material surface (e.g., molecular adsorption and associated dipole moments) enter this expression only through their effect on the work function as a transmission barrier which determines the width of the kinetic energy distribution and consequently the yield.
On the other hand, surface effects, such as the vectorial photo-effect \cite{broudy_vectorial_photoeffect}, are highly sensitive to polarization and incidence angle of the light which enables their experimental separation from bulk emission. Surface effects have been found to increase the yield up to $20$\,\% at incidence angles of around $60^\circ$ with respect to the surface normal \cite{hechenblaikner_photoemission}. The total quantum yield of an arbitrary surface is illustrated by the number of emitted electrons in Figure~\ref{fig:principle_discharging}.

\subsection{Photoemission Sensitivity}
\label{subsec:contamination}
The quantum yield of gold surfaces under UV illumination with energy close to the surface work function is heavily influenced by gross contamination from packaging, shipping, and handling, as well as by contaminant substances adsorbed from the atmosphere, in particular water and hydrocarbons. Although gross contamination is removed from the discharge relevant surfaces by an initial (plasma) cleaning step right before the inertial sensor assembly, integration, and test (AIT) procedure, it can not be avoided that these surfaces will be exposed to air before they are brought into vacuum with a typical initial pressure of $10^{-7}$\,mbar.

As shown in \cite{hechenblaikner_photoemission}, the emission efficiency of cleaned (Ar-Ion sputtered) gold surfaces initially increases due to the air exposure, before diminishing below the detection limit of the used measurement apparatus on a timescale of several hours. Outbaking at about $125\,^{\circ}\mathrm{C}$, restored the emissivity and also reduced the quantum yield variation below a factor of approximately three (calculated as the ratio between the minimum and maximum quantum yield from $8$ independent sample measurements). 

If the gold surfaces are contaminated, their work function changes to $\Phi'=\Phi-\Delta\Phi$ due to the adsorption of the contaminants, where $\Delta\Phi$ is typically between $0.3$\,V and $0.8$\,V \cite{hechenblaikner_photoemission}. Assuming the photo-current obtained from bulk emission scales according to Eq.~\ref{eg:fowler}, this implies that a work function variation $\Delta\Phi$ will change the photo-current dramatically if $h\nu$ is very close to $\Phi$, but only moderately if $h\nu\gg\Phi$. In order to reduce the sensitivity to surface contaminations, it is therefore desirable to have $h\nu\gg\Phi$ so that surface contaminations have only a moderate impact on the total photo-current. Work functions for air-exposed gold surfaces are typically in the order of $\approx4.1$\,eV \cite{feuerbacher_1972experimental}\cite{zeting_cosine_distribution}, which we also found from initial measurements with minimal water adsorption \cite{hechenblaikner_photoemission}.

\subsection{UV Light Actuation Strategies}
\label{subsec_designconcepts}
As indicated in Figure~\ref{fig:principle_discharging}, the electric fields strongly affect the electron trajectories and, obviously, they can be used to enhance or attenuate the transition of electrons. In combination with the selected UV light injection into the inertial sensor, this effect can be utilized for test mass discharging purposes. In this context, one may distinguish two different UV light actuation concepts: 

\begin{enumerate}
	\item Switch on the UV light only at periods in time when the present voltages support the desired electron flow and consequently attenuate (or suppress) the unwanted electron flow. Ideally, for positive test mass discharge rates, the electron transition ratios from the TM to the EH surfaces are equal to one (i.e., all electrons make the transition) and those from the EH to the TM surfaces are equal to zero (i.e., no electron makes the transition), and vice versa for negative discharge rates. The concept works although the test mass and the electrode housing surfaces are illuminated at the same time (as long as both sides emit photo-electrons).
	\item Switch on the UV light constantly, irrespective of any applied voltages. Ideally, a positive discharge rate is obtained when only the TM is illuminated; a negative discharge rate for pure EH illumination. Usually, both sides are illuminated due to reflections; therefore, bipolar discharging is obtained by ``adjusting'' a specific imbalance of EH and TM illumination (achieved by different UVLFs which either point towards the EH or towards the TM). The transition ratios are somewhere between $0$ and $1$, depending on the (time-variant) voltages applied to the electrodes.
\end{enumerate}

The first concept is referred to as ``synchronized AC charge control'' in the following, and the latter as ``unsynchronized DC charge control''. In this context, AC and DC refer to the on/off switching frequency of the UV light with respect to the applied voltages. AC switching means to 
turn on the light only for a short fraction of time (compared to the period of the alternating voltages applied to the electrodes). DC switching means that the light is turned on much longer, such that it can be considered constant w.r.t. the alternating voltages. Ideally, the second concept generates bipolar test mass discharge rates without application of any voltages; the averaged transition ratios can be increased or decreased through application of DC bias voltages. The first concept requires the application of alternating (e.g., sinusoidal) voltages to obtain bipolar discharge rates.

\section{Discharge Model}
\label{discharge_model}
The following sections~\ref{discharge_model_raytracer} and \ref{discharge_model_electrontracer} describe comprehensive numerical tools, which have been developed to model and analyze UV discharge systems. Dedicated hardware measurements can be taken into account as model input parameters. Section~\ref{subsec:simplifiedmodel} introduces an analytical discharge model which makes use of outputs from the developed analysis tools.

\subsection{Ray Tracing Tool}
\label{discharge_model_raytracer}
\begin{figure}[!t]
\centering
\includegraphics[width=0.46\textwidth]{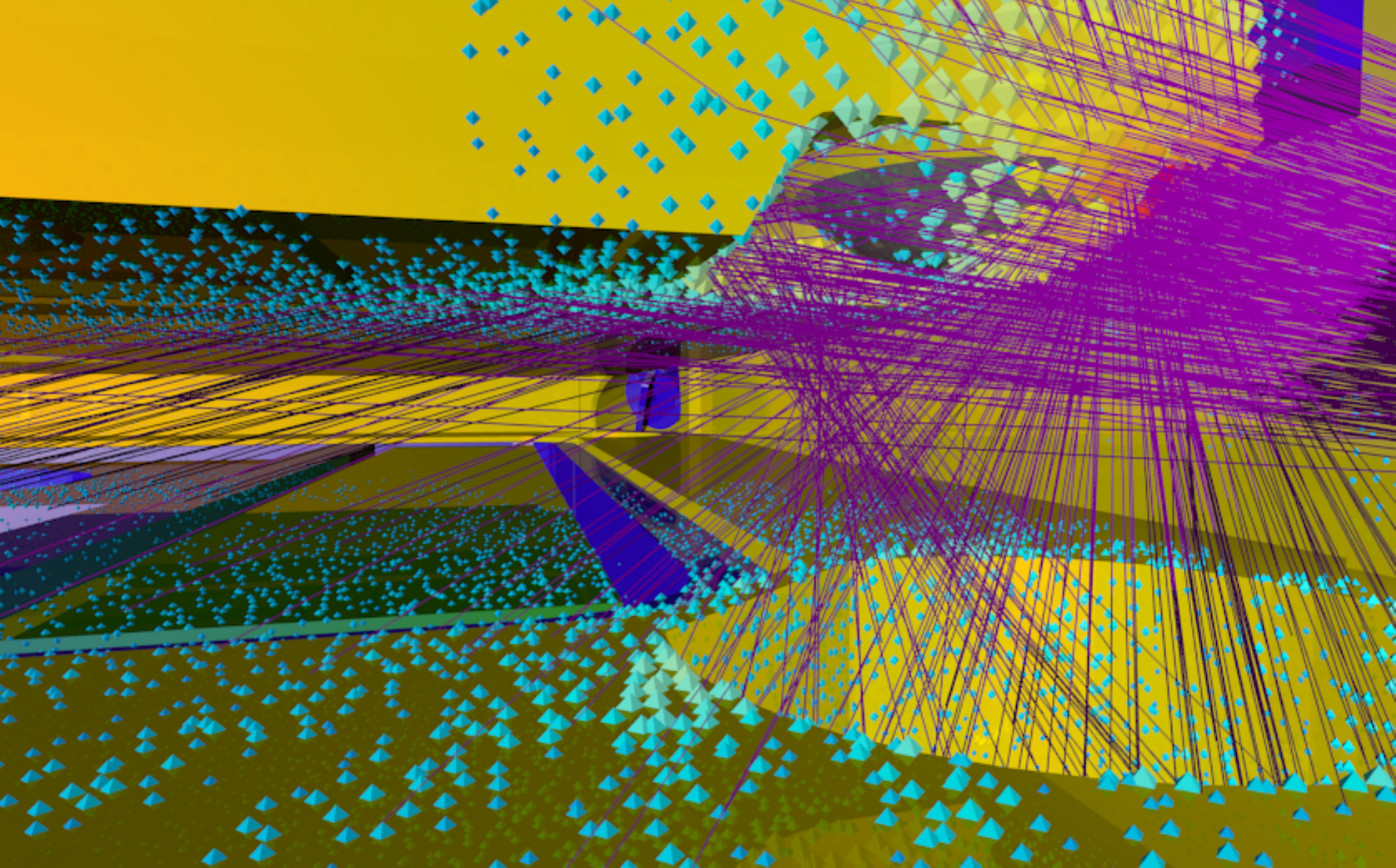}
\caption{Visualization of the three-dimensional geometry model of the LPF inertial sensor as used for ray tracing simulations. The scene shows a test mass corner with the spherical launch lock interface (upper part of the figure) inside the housing structure with the electrodes and caging finger holes (bottom part). The scene is illuminated by the UVLF pointing towards the test mass (no. 1 in Fig.~\ref{fig:LPF_IS_Explosion}). Also visualized are some UV light rays and the absorbed light power at the impact positions.}
\label{fig:geometry_and_impacts}
\end{figure}

In order to predict photo-currents from individual inertial sensor surfaces, the absorbed UV light power of these surfaces must be known. The spatial light absorption within an integrated sensor cannot be measured directly. For simple geometries, the light distribution may be calculated using analytical methods; however, this is not possible for the complex inertial sensor geometry. Therefore, a ray tracing tool has been developed to calculate and visualize the absorbed light power of arbitrary sensor surfaces. The tool approximates the exact solution by tracing a large number of generated rays through the scene. When a ray hits a surface of the geometry, it is partly absorbed and partly reflected, according to the material's reflection and scattering properties. The reflected ray is further traced through the scene until its remaining power is negligible. Figure~\ref{fig:geometry_and_impacts} shows the gap between test mass and electrode housing, generated by the three-dimensional geometry model of the reference inertial sensor (see Appendix~\ref{subsec:ISdescription}). Also the ray-tracing results are visualized by a limited number of UV light rays together with the absorbed light power at the corresponding light impact positions.
\\

\subsubsection{Ray tracer inputs and measurements}
The scene geometry is directly imported from the CAD model of the inertial sensor flight hardware. Hence, all mechanical design details (e.g., specific electrodes, caging finger holes, test mass corners, etc.) are completely considered in the three-dimensional geometry model. The materials of all surfaces can be assigned since they define the physical properties as needed by the ray tracing simulation. 

\begin{figure}[!t]
\centering
\includegraphics[width=0.46\textwidth]{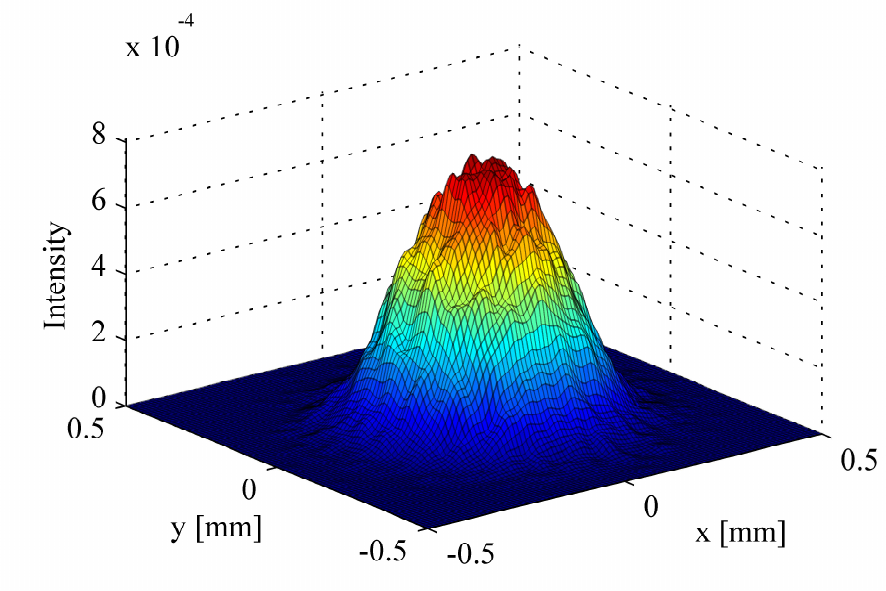}
\caption{Processed measurement data of a typical two-dimensional UV light intensity distribution as emitted from an UVLF with a $1$\,mm multimode fiber inside. The raw data used to generate the distribution has been provided by Imperial College London, the manufacturer of the LPF UV light source hardware.}
\label{fig:spectrum_light_distribution}
\end{figure}

A light source model generates the light rays as emitted from the tips of the UVLFs. The rays are randomly generated such that the simulated light source characteristics corresponds to the wavelength spectrum and the spatial intensity distribution of the real light source. As an example, the processed measurement data of a typical light intensity distribution from an UVLF tip is shown in Figure~\ref{fig:spectrum_light_distribution}. The distribution is one of the input parameters needed by the random number generator of the light source model.

\begin{figure}[!t]
\centering
\includegraphics[width=0.46\textwidth]{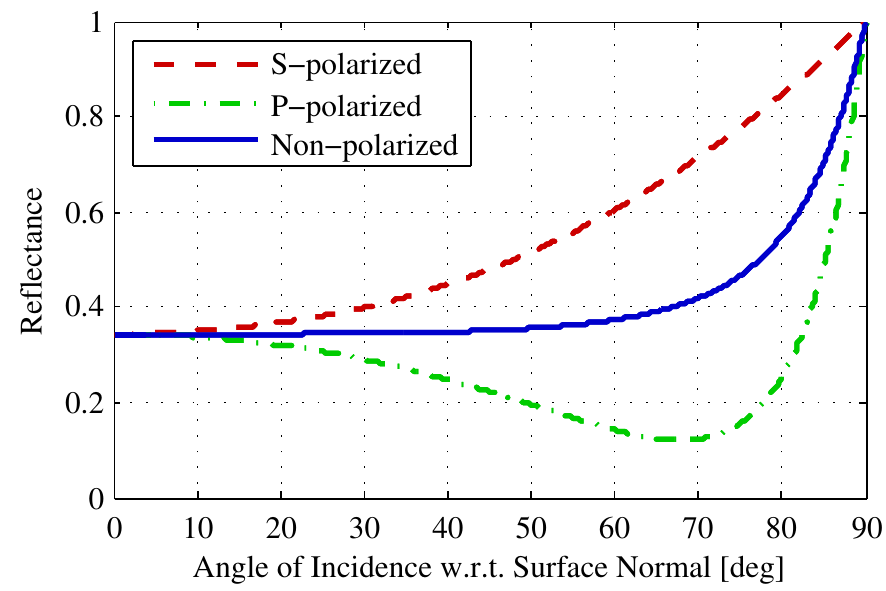}
\caption{Reflection curve of a typical electrode gold surface ($R_a<10$\,nm) versus UV light illumination with $254$\,nm wavelength at different incident angles. The measurement of the non-polarized reflection curve has been performed by ESA on a representative surface sample of an inertial sensor electrode.}
\label{fig:reflection curve}
\end{figure}

Each time a light ray hits an inertial sensor surface, some of its power is absorbed by the surface and the rest is reflected. The amount of reflected and absorbed light depends on the material, the angle of incidence (defined with respect to the surface normal), the wavelength and the polarization of the light. The reflectance of a typical inertial sensor gold surface versus angle of UV light incidence is shown in Figure~\ref{fig:reflection curve}. 

For the scattering of the reflected light, a simple rule of thumb is that specular reflection (light incident angle = reflection angle) can be assumed if the mean surface roughness $R_a$ is smaller than $1/20$ of the light wavelength $\lambda$. Otherwise, diffuse scattering has to be considered to obtain a realistic light propagation after reflection from ``rough'' surfaces. The amount and distribution of UV light scattering from rough sensor surfaces is characterized by the bidirectional reflectance distribution function (BRDF). The BRDF is a 5-dimensional function and describes the ratio of the reflected radiance P$_s$ exiting from the surface in a particular direction, to the directed irradiance P$_i$ incident on the surface. It is defined as
\begin{equation}
\label{eg:brdf}
\text{BRDF}(\theta_i,\phi_i,\theta_s,\phi_s,\lambda)\,=\,\frac{P_s}{\Omega_s P_i \cos(\theta_s)}  
\end{equation}
where the subscripts $i$ and $s$ mean incident and scattered. $\Omega_s$ is the solid angle at which the scattered radiance is measured and depends on the size of the detector; the term $\cos(\theta_s)$ is a correction factor to adjust the illuminated area to its apparent size when viewed from the scatter direction. The angles $\phi_s$ and $\theta_s$ are the azimuth and the zenith angle with respect to the surface normal. They describe the hemisphere above the surface in which the radiance P$_s$ of the reflected beam can be characterized. The BRDF further depends on the azimuth angle $\phi_i$ of the incident beam (only needed for anisotropic materials), its zenith angle $\theta_i$ (corresponding to the incidence angle w.r.t the surface normal), and on the light wavelength $\lambda$. 
In the ray tracer, BRDF functions are represented by at set of measurements, obtained from a robot-based gonioreflectometer\footnotemark\footnotetext{In this context, a gonioreflectometer is a device which allows the precise control of the angles of the incident and reflected light beams in a reflection measurement.} apparatus \cite{hope_brdf_measurements} for various light incident angles. 
Practically, the BRDF can only be measured for a limited combination of the five relevant parameters; therefore, a weighted linear interpolation method is used to derive data sets for all possible light incident angles and wavelengths from the available measurement sets. Subsequently, for each incident angle, a two dimensional light distribution probability density function (PDF) can be calculated from the interpolated measurement data. The direction of the reflected light ray is then randomly generated according to this PDF.\\

\subsubsection{Ray tracer outputs}
The output of the ray tracer is a list of all occured ray impacts on the inertial sensor surfaces. For each impact, the absorbed light power, the impact position, the angle of incidence with respect to the surface normal, and the light wavelength of the incident ray are stored. The integrated power absorption of all impacts on an arbitrary surface $i$ (e.g., an actuation/sensing electrode, the test mass, etc.) is denoted \emph{illumination ratio} $\rho_i$, where $i$ is the surface identification number. The illumination ratios of all surfaces can be derived by post-processing of the ray-tracing data and correspond to a fraction of the total injected light power such that $\sum_i\,\rho_i\,=1$. The accuracy of the solution increases with the number of generated rays and converges towards the analytical solution for an infinite number of rays. For a typical inertial sensor illumination scenario, convergence of relevant surface illumination ratios (i.e., the numerical variation of the calculated values is less than $1$\,\%; looking only at surfaces where $\rho_i > 1.0$\,\%) is reached by tracing $\approx10^6$ rays.

\subsection{Electron Tracing Tool}
\label{discharge_model_electrontracer}
In order to predict discharge rates between inertial sensor surfaces, the probability of released photoelectrons to travel from surface $i$ to another surface $j$ must be known. The prediction of individual electron trajectories is sophisticated, especially for complex geometries (e.g., like the illuminated caging finger holes and the test mass corner spheres) and in regions with inhomogeneous electric fields. Note that the LISA Pathfinder discharge system design strongly illuminates exactly such regions (see Figure~\ref{fig:geometry_and_impacts}).

For this reason, an electron tracing tool has been developed to propagate the trajectories of individual electrons from their origins to their absorption points on the receiving surface. The complex and time-variant electric field in the gap between the surfaces is considered when the electron trajectories are simulated. The goal is to obtain the transition ratios between all pairs of surfaces for a specific interval of time.\\

\subsubsection{Electron tracer inputs and measurements}
\label{electrontracer_inputsandmeas}
The electron trajectories are simulated considering the following inputs:
\begin{itemize}
		\item CAD model of the inertial sensor geometry
		\item Absorbed light power and impact position as simulated by the ray-tracing tool
		\item Measured kinetic energy probability density function of the photo-electrons
		\item Angular distribution of the photo-electrons
		\item Time-variant electric fields due to applied voltages and test mass charge
		\item Time of electron release w.r.t. the applied voltages
		\item Measured quantum yields of the emitting surfaces
\end{itemize}
For each electron, an initial condition (time, release position, and velocity vector) is needed. Based on these initial conditions and the instantaneous electric field, electrons are traced through the geometry by solving their equations of motion. The tracing is stopped when the electron hits its receiving surface.

The release position of the electrons is obtained from the light impact positions as previously calculated by the ray tracing tool. 

\begin{figure}[!t]
\centering
\includegraphics[width=0.46\textwidth]{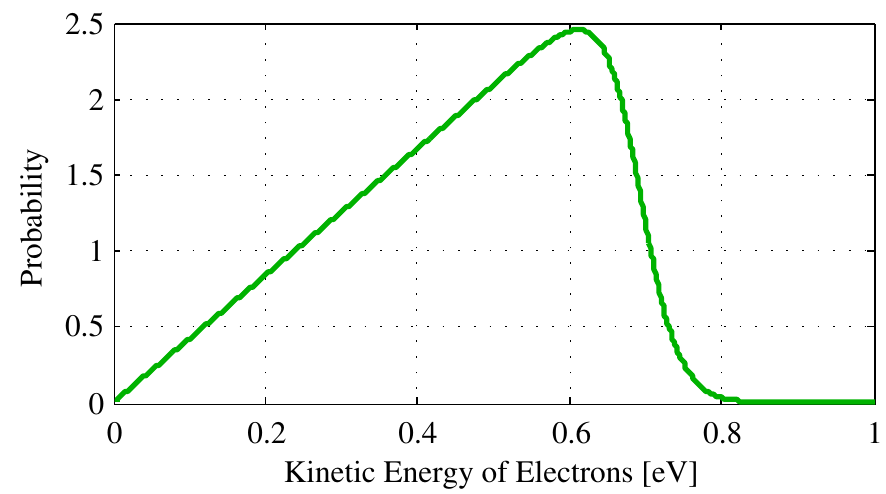}
\caption{Kinetic energy distribution of electrons emitted from air-exposed gold surfaces under UV illumination with $253.6$\,nm wavelength at $\approx300$\,K.}
\label{fig:duBridge}
\end{figure}

The initial velocity is computed from the kinetic energy distribution of the photo-electrons. A typical distribution, used as electron tracer parameterization for the gold surfaces, is shown in Figure~\ref{fig:duBridge}.
It closely resembles the distribution for the total energy according to DuBridge \cite{dubridge_theory_energy_distribution_photoelectrons} for emission from a free electron gas at $300$\,K; which has been found by detailed analysis of measurement data obtained from specifically conducted sample measurement campaigns \cite{hechenblaikner_photoemission}. The analysis considers the electrode geometry of the measurement apparatus, the impact of dipolar adsorbants, and the effect of disturbing electric fields that have been observed during the measurements conducted at the German Aerospace Center (DLR). 
For each electron to be traced, a random initial kinetic energy $\text{E}_\text{kin}(X_1)$ is generated by mapping a pseudo random number $X_1$ between $0$ and $1$ (generated from a uniform distribution) to the input range of an inverse cumulative distribution function. The cumulative distribution is derived from the measured probability density function (e.g., as the one shown in Figure~\ref{fig:duBridge}), which is assigned to the corresponding sensor surfaces. The initial electron velocity is given as  
\begin{equation}
\label{eg:initial_velocity}
\left|v_0\right|\,=\,\left|\sqrt{\frac{2\cdot\text{E}_\text{kin}(X_1)}{m_{e}}}\right|,  
\end{equation}
where $m_e$ is the electron mass at zero velocity.

\begin{figure}[!t]
\centering
\includegraphics[width=0.35\textwidth]{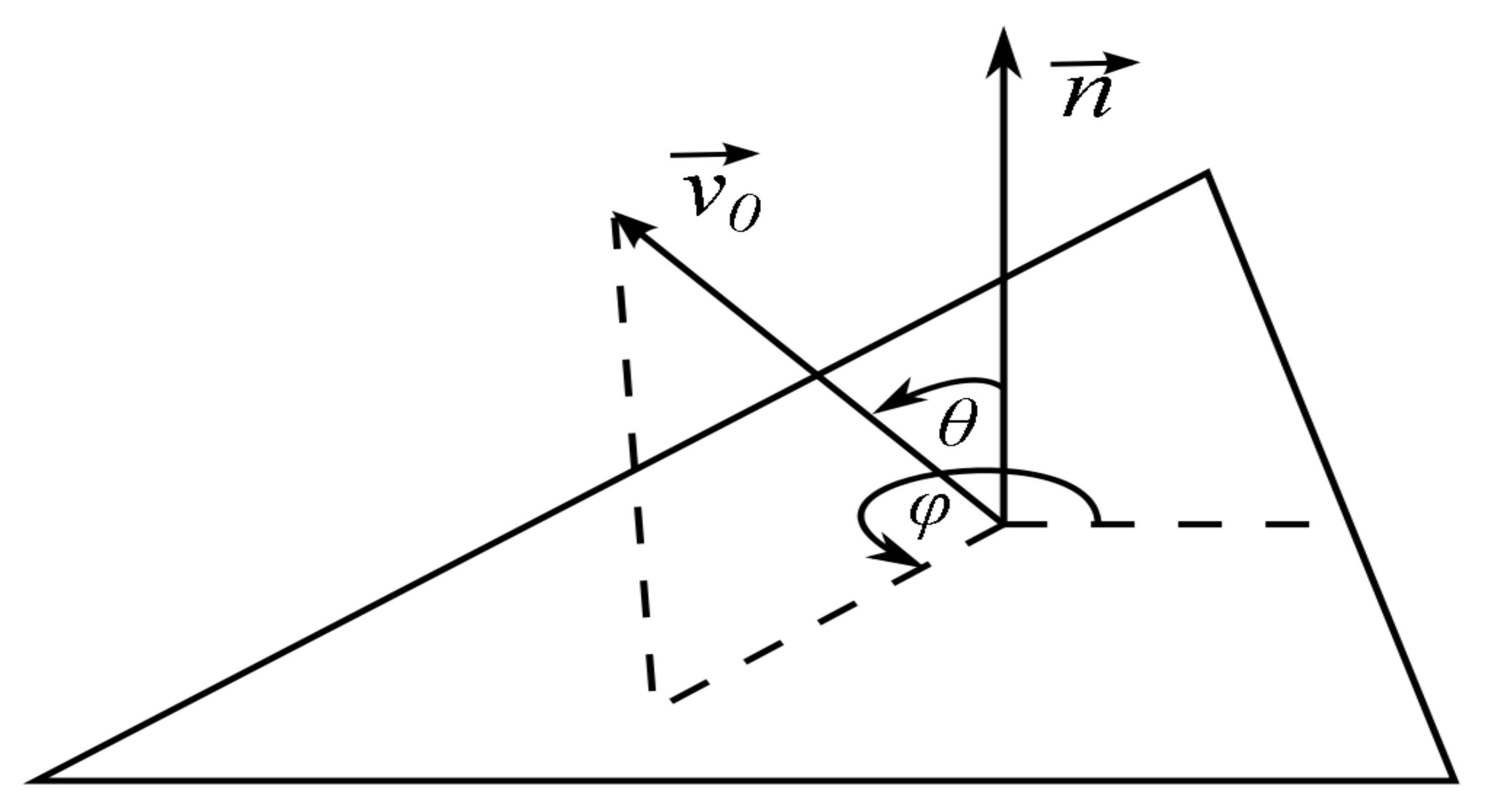}
\caption{Illustration of an initial velocity vector $\vec{v_0}$ of an emitted electron w.r.t. the surface normal $\vec{n}$.}
\label{fig:release_vector}
\end{figure}

The release directions of photo-electrons follow a cosine-distribution \cite{zeting_cosine_distribution}\cite{hechenblaikner_photoemission}. Figure~\ref{fig:release_vector} illustrates the initial velocity vector $\vec{v_0}$ of an electron with respect to the surface normal vector $\vec{n}$. The direction is defined by the angles $\theta$ and $\varphi$ according to:
\begin{eqnarray}
 \label{eq:cosine_distribution} 
\theta   &=& \arccos\left( \sqrt{1-X_2}\right) \nonumber \\
\varphi  &=& 2\pi\cdot X_3,  
\end{eqnarray}
where $X_2$ and $X_3$ are uniformly distributed pseudo random numbers between $0$ and $1$. The velocity vectors are distributed on a half sphere with the emitted electron in the center.

The force acting on an electron in a time-variant electric field is given as
\begin{equation}
\vec{F}\left(\vec{r},t\right)\,=\,e \cdot \vec{E}\left(\vec{r},t\right)\,=\, m_e\cdot \ddot{\vec{r}}\left(\texttt{}\vec{r},t\right).
\label{eq:electron_eom1}
\end{equation}
In Eq.~\ref{eq:electron_eom1}, $e$ is the electron charge and $\vec{E}\left(\vec{r},t\right)$ the strength of the electric field at electron position $\vec{r}$ and time $t$. The acceleration $\ddot{\vec{r}}\left(\vec{r},t\right)$ on the electron can be written as a second-order differential equation:
\begin{equation}
\label{eq:electron_eom2}
\ddot{\vec{r}}\left(\vec{r},t\right)\,=\,\frac{e}{m_e} \cdot \vec{E}\left(\vec{r},t\right).
\end{equation}
The electron trajectory is the solution of Eq.~\ref{eq:electron_eom2}.  
Due to the inhomogeneous electric field and the complex geometry of the inertial sensor, it is usually not possible to calculate an analytical solution. Therefore, Eq.~\ref{eq:electron_eom2} is solved numerically, using the explicit 4th/5th order Runge-Kutta one-step solver. The electric field at time $t$ can be assumed \emph{static} during the computation of individual electron trajectories, since their transit time in typical (millimeter-sized) sensor gaps is a few nanoseconds and the periods of the applied sinusoidal voltages (see Appendix~\ref{subsec:ISdescription}) are much larger.

The local strength $\vec{E}\left(\vec{r},t\right)$ of the time-variant electric field inside the inertial sensor is computed using a commercially available finite-element software, where the same CAD model as used for the ray tracing has been imported for consistent geometry definition. It would require many calculations to obtain the electric fields inside the sensor for each possible combination of time-variant electrode voltages and test mass potentials that might occur during operation. Thus, the following approach is adopted: 

The influence of each individual body (electrodes, housing structure, test mass) on the total electric field is computed separately by the finite-element software, for a constant potential of $1$\,V, while all other bodies are set to zero potential. With the obtained sets of electric fields, the total electric field can be reconstructed during electron tracing at each time step and for every possible combination of electrode potentials. This is achieved by superposition of the individual $1$\,V-field solutions, where each field is scaled according to the currently applied electrode voltages and the resulting test mass potential. 

The finite-element software provides the $1$\,V-field solutions as locally discretized vector fields (the meshing in the finite-element calculation has been selected to limit the energy error of each field solution below $0.1$\,\%). Therefore, during the numerical integration of Eq.~\ref{eq:electron_eom2}, each $1$\,V-field has to be interpolated to obtain the field strength at the current electron position. This is done by means of a vectorial interpolation method based on spatial Delaunay segmentation \cite{www_qhull} and barycentric-rating.
Hence, the method allows to trace electrons at arbitrary instants of time, covering all possible electric field situations within the sensor.

The photoemission of an arbitrary surface is modeled as the product of the quantum yield from bulk emission $\xi$ (defined as the number of emitted electrons per absorbed photons) and the quantum yield gain $g$, which depends on the angle of light incidence. The gain models the influence of the vectorial photo-effect which might be caused by surface effects. The value of the gain is normalized to $1$ at normal incidence and usually grows for increasing incidence angles \cite{hechenblaikner_photoemission}.\\ 

\subsubsection{Electron tracer outputs}
The main output are electron transition ratios between arbitrary sensor surfaces. The transition ratio $f_{i\rightarrow j}(t)$ from surface $i$ to another surface $j$ is computed as the weighted sum over all traced electrons that travel from surface $i$ to surface $j$, divided by the weighted sum over all released electrons from surface $i$. The weight of each electron is given by the product $\xi_m\cdot g_m \cdot \rho_m$, where $m$ is the unique identification number of an electron. In this context, $\xi_m$ is the (bulk emission) quantum yield of the emitting surface, $g_m$ the quantum yield gain, and $\rho_m$ is the fraction of absorbed light at the ray impact (and electron release) position. 

\begin{figure}[!t]
\centering
\includegraphics[width=0.46\textwidth]{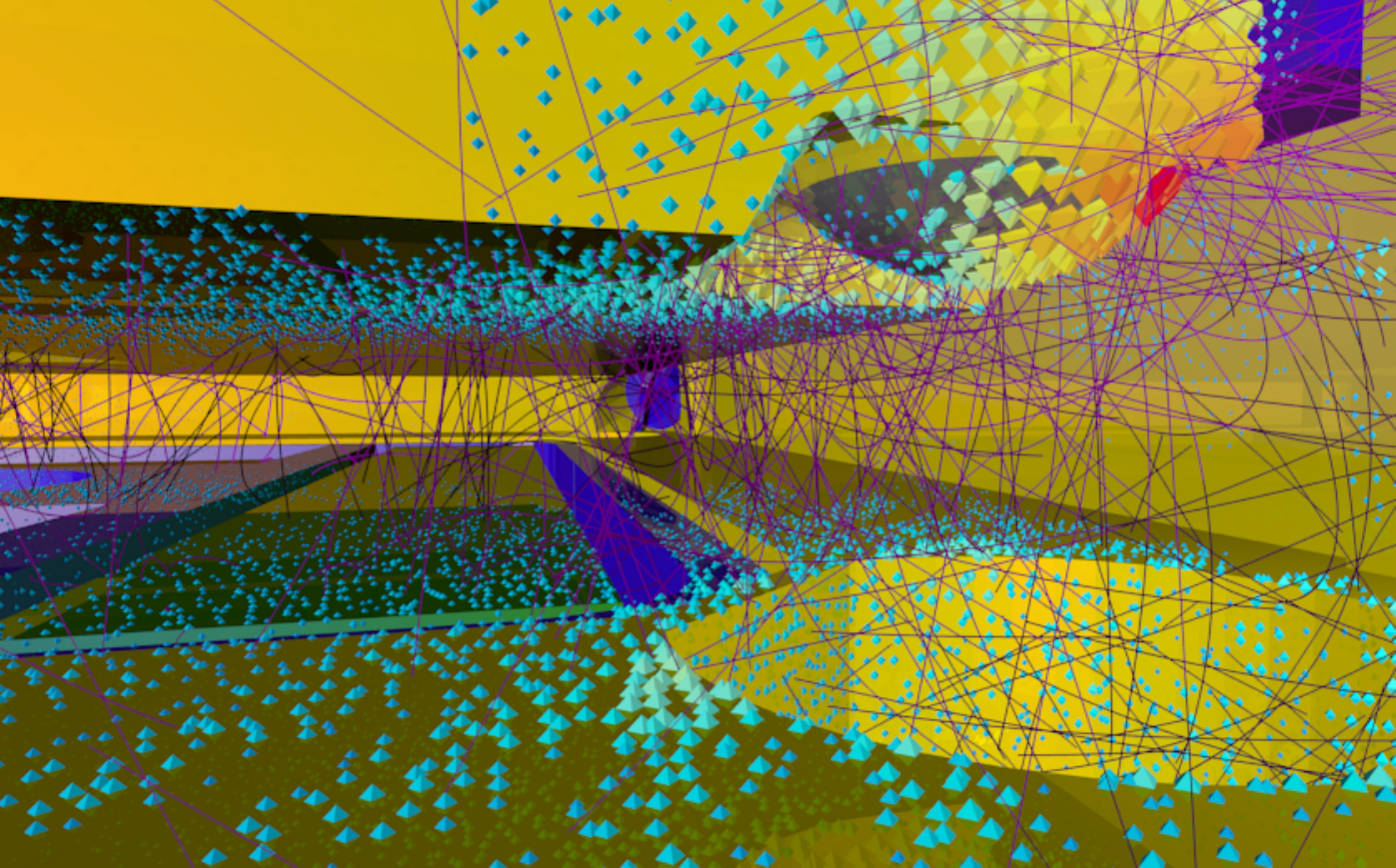}
\caption{Visualization of some electron trajectories. The trajectories are simulated for the LISA Pathfinder discharge system, where the test mass is illuminated (as shown in Figure~\ref{fig:geometry_and_impacts}) and a typical high frequency voltage for electrostatic sensing is present (see Eq.~\ref{eq:injectionvoltage} in Appendix~\ref{subsec:ISdescription}). The trajectories are valid only for a snapshot in time and notably vary due to the time-variant electric field.}
\label{fig:electron_trajectories}
\end{figure}

Due to the applied AC voltages, the transition ratios are functions of time. Usually, their values significantly change with the amplitude and orientation of the electric field. Theoretically, the value at a specific point in time $t_x$ is simply obtained by considering only electrons that have been emitted exactly at time $t_x$. Practically, the transition ratios at time $t_x$ are obtained by considering all simulated electrons in the small interval $\Delta t=[t_x-\tau<t_x<t_x+\tau]$. The computed trajectories for a certain time interval $\Delta t$ can be visualized, as shown in Figure~\ref{fig:electron_trajectories}.

\subsection{Analytical Discharge Model}
\label{subsec:simplifiedmodel}
The simulated electron trajectories can be used to set up an analytical model for the instantaneous charge rate $\dot{Q}_{i\rightarrow j}(\Delta t)$ from an arbitrary surface $i$ to another surface $j$:
\begin{equation}
\label{eq:simplified_discharge_model1}
\dot{Q}_{i\rightarrow j}(\Delta t) = \Bigg(\sum_{m=1}^M \xi_m \cdot g_m \cdot \rho_m \cdot \delta_{i\rightarrow j}(\Delta t) \Bigg) \cdot I_\text{UV}(\Delta t).
\end{equation}
In Eq.~\ref{eq:simplified_discharge_model1}, $M$ is the total number of traced electrons within the
small time interval $\Delta t$ and $I_\text{UV}(\Delta t)$ is the injected radiant UV light power. The function $\delta_{i\rightarrow j}(\Delta t)$ becomes one when the electron $m$ makes a transition from surface $i$ to $j$ within $\Delta t$ (as can be identified from its computed trajectory); otherwise it is zero. The transition ratio $f_{i\rightarrow j}(\Delta t)$ from surface $i$ to surface $j$ at time $\Delta t$ is given as:
\begin{equation}
\label{eq:transition_ratios}
f_{i\rightarrow j}(\Delta t) = \frac{\sum_{m=1}^M \xi_m \cdot g_m \cdot \rho_m \cdot \delta_{i\rightarrow j}(\Delta t)}{\sum_{m=1}^M \xi_m \cdot g_m \cdot \rho_m \cdot \delta_{i}(\Delta t)},
\end{equation}
where $\delta_{i}(\Delta t)$ becomes one when electron $m$ is released from surface $i$ in the interval $\Delta t$. The instantaneous \emph{net} charge rate of an arbitrary surface $i$ is given as:
\begin{equation}
\label{eq:simplified_discharge_model2}
\dot{Q}_{i}(\Delta t) = \sum_{j=1}^N \Big(\dot{Q}_{i\rightarrow j}(\Delta t) - \dot{Q}_{j\rightarrow i}(\Delta t) \Big). 
\end{equation}
$N$ is the total number of modeled inertial sensor surfaces. 

In order to quantify the robustness of a discharge system, not only the instantaneous discharge rates are of interest, but also the \emph{averaged} discharge rates within an arbitrary time span $T>\Delta t$. The averaged electron flow from surface $i$ to surface $j$ is obtained when Eq.~\ref{eq:simplified_discharge_model1} is evaluated for all electrons generated within $T$. Moreover, the \emph{overall} averaged electron flow $\bar{\dot{Q}}_{\text{TM} \rightarrow \text{EH}}(T)$ from the TM to the EH is given by the sum of all individual electron flows from the test mass surfaces ($\forall\,i \in \text{TM}$) to the EH surfaces ($\forall\,j \in \text{EH}$). Each electron flow from TM surface $i$ to EH surface $j$ is computed according to Eq.~\ref{eq:simplified_discharge_model1} for all electrons generated within $T$. The averaged electron flow $\bar{\dot{Q}}_{\text{EH} \rightarrow \text{TM}}(T)$ from the EH to the TM is derived similarly.

The individual (averaged) electron flows between test mass and electrode housing surfaces can be used to quantify the robustness $\Delta \dot{Q}$ of a discharge system design to obtain bipolar test mass discharge rates (e.g., despite the presence of quantum yield variations between illuminated surfaces). For example, a positive discharge rate $\dot{Q}_\text{TM}^+$ is obtained when:
\begin{equation}
\label{eq:general_robustness}
\Delta \dot{Q}^+ = \frac{\bar{\dot{Q}}_{\text{TM} \rightarrow \text{EH}}(T)}{\bar{\dot{Q}}_{\text{EH} \rightarrow \text{TM}}(T)}>1.
\end{equation}
The actual value of the ratio $\Delta \dot{Q}^+$ quantifies the robustness to achieve positive discharge rates; a value $\Delta \dot{Q}^+ \gg 1$ indicates that a positive discharge rate can be safely achieved, although the quantum yields of the EH surfaces are larger than that of the TM surfaces. Negative discharge rates are obtained when the reciprocal value of Eq.~\ref{eq:general_robustness} is larger than $1$. The desired robustness gain for a negative discharge rate is  $\Delta \dot{Q}^- \gg 1$; hence, it can be realized even when the TM quantum yields are larger than that of the EH.

\section{Analysis of the LISA Pathfinder UV Light Discharge System}
\label{sec:desinganalysis}
In this section, the discharge toolbox presented in Section~\ref{discharge_model} is used to analyze the baseline LISA Pathfinder discharge system \cite{sumner_description_LISA_chargingdischarging}\cite{shaul_charge_management_LISA_and_LPF} as well as of a design modification that has been proposed to increase the system robustness.  


\subsection{Baseline LISA Pathfinder Discharge System}
\label{subsec:LPFdischargesystem}
The UV light is generated by Hg-discharge lamps of which wavelengths shorter than $230$\,nm are removed by a filter so that only light of $253.6$\,nm wavelength ($4.89$\,eV) is transmitted. 
The radiant power can be controlled by setting one of $256$ different commands, at a maximum rate of $1$\,Hz. As detailed in Appendix~\ref{subsec:ISdescription}, the frequencies of the sinusoidal voltages applied to the different electrodes range between $60$\,Hz and $100$\,kHz. Hence, the LPF discharge principle is classified ``unsynchronized DC charge control'', according to the definition in Section~\ref{subsec_designconcepts}. Measured calibration tables, relating the command setting and the radiant power $I_\text{UV}$ emitted at the tip of the UV light feedthroughs, are available. The dynamical range (defined as the ratio of maximum and minimum stable photon output) is approximately $100$.

The feedthrough pointing towards the TM (denoted UVLF~$1$ in Figure~\ref{fig:LPF_IS_Explosion} of Appendix \ref{subsec:ISdescription}) is normally used to generate positive discharge rates; the feedthroughs pointing towards the housing structure (denoted UVLF~$2$ and $3$ in Figure~\ref{fig:LPF_IS_Explosion}) are used to generate negative discharge rates.

The measured light distribution emitted from a feedthrough tip has been processed in order to obtain a two-dimensional distribution function (similar to the one shown in Figure~\ref{fig:spectrum_light_distribution}). The processed light intensity distribution is used as input parameter for the light source model of the developed ray tracing tool. 
The LPF inertial sensor geometry is directly obtained from the CAD model of the flight sensor and the materials with their corresponding reflection and scattering properties are assigned to the individual sensor surfaces. For the UV wavelength of $253.6$\,nm, the critical surface roughness to assume specular reflection is $\lambda/20\approx13$\,nm. Therefore, non-specular scattering effects have to be considered\footnotemark\footnotetext{The presented analysis has been performed without characterization of rough sensor surfaces through BRDF measurements; however, BRDF measurements are currently being performed at PTB Braunschweig and will be used to refine the simulations.} when the mean surface roughness $R_\text{a}\gg13$\,nm (which is the case for the electrode housing structure and the test mass corner spheres). 

The initial velocities of the photo-electrons are obtained from the kinetic energy distribution measurements described in Section~\ref{electrontracer_inputsandmeas}. In particular, for the various gold coated sensor surfaces, the distribution shown in Figure~\ref{fig:duBridge} is used as input parameterization for the electron tracing tool. 

\begin{figure}[!t]
\centering
\includegraphics[width=0.46\textwidth]{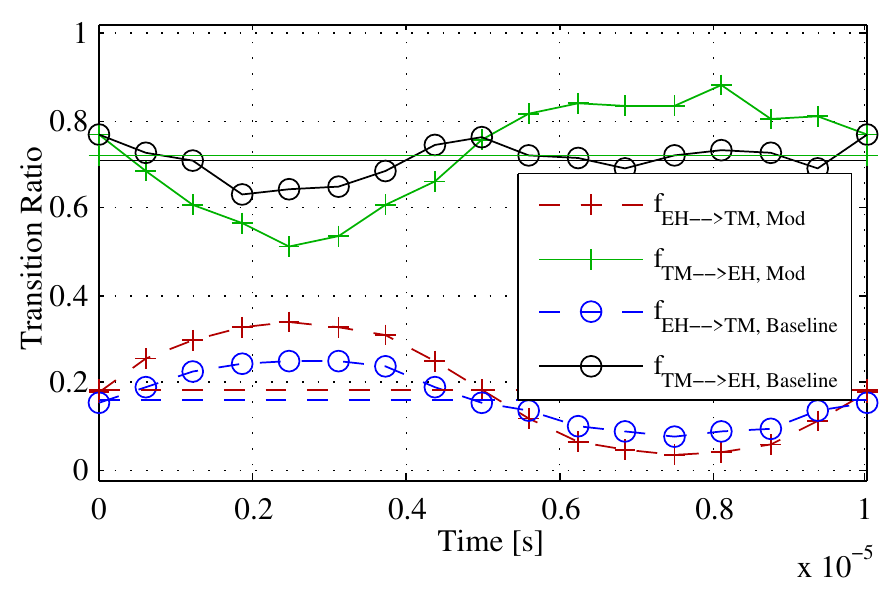}
\caption{Time-variant and averaged electron transition ratios for negative TM discharge rates, plotted over $10\,\mathrm{\mu}$s (corresponding to one period of the injection voltage). The EH and TM illumination ratios obtained with UVLF~$2$ are: $\rho_\text{EH}=88.4$\,\% and $\rho_\text{TM}=11.6$\,\% (baseline design); $\rho_\text{EH}=98.4$\,\% and $\rho_\text{TM}=1.6$\,\% (design modification).}
\label{fig:LPF_Transition_ratios}
\end{figure}

From the ray tracing and subsequent electron tracing simulations, it became clear that a large amount of photo-electrons do not contribute to the discharge rates as desired. Especially for the negative discharge rates, obtained by illumination of the electrode housing, only few of the electrons emitted inside the strongly illuminated caging finger holes reach the test mass. Moreover, a considerable part of the UV light is reflected from the EH to the TM ($\rho_\text{TM}=11.6$\,\%), where $72$\,\% of the emitted electrons make the transition to the housing ($f_{\text{TM}\rightarrow \text{EH}}(t)=0.72$). The calculated transition ratios $f_{\text{EH}\rightarrow \text{TM}}(t)$ and $f_{\text{TM}\rightarrow \text{EH}}(t)$ for EH illumination with UVLF~$2$ are shown in Figure~\ref{fig:LPF_Transition_ratios}, over one period of the injection voltage. The effect of the low-frequency actuation voltages is considered, since the shown values at each point are mean values which have been calculated from $10^5$ injection voltage periods within one second.

The predicted robustness gains $\Delta \dot{Q}$ of the baseline LPF discharge system to achieve bipolar test mass discharge rates are reported in Table~\ref{table_robustness_LPF}. Assuming the quantum yields of all \emph{not} gold coated surfaces to be zero (caging fingers, TM corner spheres), $\Delta \dot{Q}$ directly shows the robustness against quantum yield imbalances of the EH and TM gold coatings. 
Thus, to obtain negative discharge rates, the quantum yield of the TM must not exceed the EH yield by the factors reported for $\Delta \dot{Q}^-$. Positive discharge rates are obtained only when the EH yield does not exceed the TM yield by the factors reported for $\Delta \dot{Q}^+$. The robustness values in Table~\ref{table_robustness_LPF} are reported for three different scenarios of applied voltages: ($1$) no DC bias voltages, ($2$) maximum possible DC bias voltages to support positive test mass discharge rates, ($3$) maximum DC bias voltages to support negative discharge rates. The effect of the injection and actuation voltages on the averaged transition ratios are considered in each scenario.

The quantum yields of representative, air-contaminated gold coated samples have been observed to vary by approximately a factor of $3$ when the sample surfaces are controlled according to a flight-model AIT process (uncontrolled surface samples have been found to vary by a factor of $70$). Therefore, as evident from Table~\ref{table_robustness_LPF}, the baseline design does not tolerate the expected yield variations without application of DC bias voltages, even when the surfaces have been controlled ($\Delta \dot{Q}^-=1.7$).

\begin{table}[h!]
\caption{\label{table_robustness_LPF} Predicted robustness of the LISA Pathfinder discharge system}
\begin{tabular}{@{}lcccc@{}} \toprule
\centering
           & \multicolumn{2}{c}{LPF Baseline Design} &  \multicolumn{2}{c}{LPF Modified Design}   \\ \cmidrule{2-5}
DC Voltage Scenario  &  $\Delta \dot{Q}^+$ & $\Delta \dot{Q}^-$ & $\Delta \dot{Q}^+$ & $\Delta \dot{Q}^-$\\ \midrule
($1$) No DC volt.    									& 4.3 & 1.7  &  4.8  &  16.1 \\
($2$) DC volt. $\dot{Q}_\text{TM}>0$  & 9.1 & --   & 10.6  &   --  \\
($3$) DC volt. $\dot{Q}_\text{TM}<0$  & --   & 9.8 &  --   &  50.9 \\ \bottomrule
\end{tabular}
\end{table}

\subsection{LISA Pathfinder Discharge System Modification}
\label{subsec:LPFdischargesystem_modification}
After failures to produce negative discharge rates have occurred during system level testing using an inertial sensor replica in a torsion pendulum configuration at University of Trento \cite{wass_testingUVdischargeforLPF}, the authors investigated means to increase the robustness of the baseline LISA Pathfinder discharge system design. The developed discharge toolbox has been used the analyze the robustness of acceptable design modifications. The finally proposed modifications are summarized below:

\begin{itemize}
\item \emph{Negative} discharge rates: The robustness has been improved through a modification of the light injection into the inertial sensor. For the given constraints to leave the feedthrough position and the UV light source unchanged (inertial sensor flight hardware and mechanical interfaces with other parts of the LTP structure are already built), the optimal solution was to re-direct the light into the $x$-gap on the $-z$-face of the EH, such that significantly less light is reflected onto the TM while slightly increasing the electron transition ratio $f_{\text{EH}\rightarrow \text{TM}}(t)$. The re-direction can be obtained by attaching a micro optical element to the tip of the UVLFs no.~$2$ and $3$.
\item \emph{Positive} discharge rate: Increase the gold coating of the strongly illuminated spherical caging finger interfaces at the test mass corners (see Figure~\ref{fig:geometry_and_impacts}). The spheres have been partly gold coated such that the fingers do not damage the coating but more electrons will be released from the test mass (gold work function is lower than that of the uncoated AuPt bulk material; thus, more electrons will be released from the test mass).
\end{itemize}

The optimization of the positive and negative discharge rates took ray-tracing and electron tracing simulations into account. For the negative discharge rate, the EH illumination has been increased from $88.4$\,\% to $98.4$\,\%. Hence, the undesired TM illumination has been reduced to $1.6$\,\% ($86$\,\% less than for the baseline design). The transition ratios for the modification are also shown in Figure~\ref{fig:LPF_Transition_ratios}. All time-variant voltages within one second have been considered in the same way as described in Section~\ref{subsec:LPFdischargesystem}. The predicted robustness values of the modified LPF discharge system are reported in Table~\ref{table_robustness_LPF}. The achieved robustness values satisfy the observed quantum yield variations by a factor of three, also for negative discharge rates without application of DC bias voltages ($\Delta \dot{Q}^-$ has been increased from $1.7$ to $16.1$). 

The achieved robustness gains are still not ideal, because outgassing in space might produce additional surface contaminations and therefore further increase the quantum yield imbalances. This may cause a malfunction of bipolar discharging, in particular when the application of DC bias voltages is not allowed, like for the test of continuous discharge control---the standard discharge mode during science measurements in LISA.

\section{Design and Analysis of a More Robust Discharge System}
\label{sec:LISAdesign}
Motivated by the marginal robustness of the LPF discharge system design, a more robust design is presented below. It can be directly applied to the reference inertial sensor with cubical test masses (see Appendix~\ref{subsec:ISdescription}). Therefore, it is proposed to be used for the future LISA/NGO mission, and, for any other space mission making use of similar inertial sensors.

\subsection{Performance Requirements}
\label{subsec:requirements}
In order to properly satisfy the LISA/NGO closed-loop charge control requirements for the fast discharge operation after a TM is released from the caging mechanism, and for continuous discharge operation during the science measurements, the discharge system shall be able to realize the following performance requirements.
 
Assuming the expected initial test mass charge level Q$_\text{TM}$ according to Section~\ref{subsec:problem_formulation}, a reasonable justification for the maximum required bipolar discharge rate is to discharge a test mass potential of $\pm1$\,V in one hour. With the total test mass to housing capacitance of $34.2$\,pF \cite{brandt_RevisedElectrostaticModel}, a maximum discharge rate requirement of $\pm6\cdot10^4$\,e/s is obtained.
A reasonable minimum discharge rate to fulfill the closed-loop charge control requirements is $\pm10$\,e/s. The required dynamic range (ratio between maximum and minimum discharge rate) for both positive and negative discharge rates is therefore $6\cdot10^3$\,e/s. Furthermore, any value between the minimum and maximum discharge rate shall be realized in steps of $\pm10$\,e/s (i.e., the charge rate resolution between the maximum positive and negative discharge rates corresponds to $1.2\cdot10^4$ steps).

\subsection{Design of the Robust Discharge System}
From the comprehensive discharge modeling and analysis it becomes clear that the UV light injection of the LISA Pathfinder discharge system is not optimal because various surfaces and complex geometrical features are illuminated. This includes the illumination of finger holes (where the released electrons have poor transition ratios to reach the test mass), test mass corners (spherical caging finger interfaces are not gold coated and have no well-defined surface structure after de-caging), different electrodes (various disadvantageous electric fields have to be considered), and large parts of the housing structure (which can not be biased directly through application of DC voltages). As becomes clear from Table~\ref{table_robustness_LPF}, the presented modifications of the LPF discharge system can only partially mitigate these disadvantages.

The LISA Pathfinder UV light source hardware unit imposes further constraints: Only unsynchronized DC operation is possible and the light output performance is characterized by strong temperature dependency and limited lifetime of the Hg discharge lamps. 

From this experience, a more robust discharge system has been designed with the following goals: 

\begin{itemize}
	\item Obtain sufficient robustness against quantum yield and work function variations such that bipolar test mass discharge rates can be safely achieved. 
	\item Satisfy the discharge rate performance requirements of Section~\ref{subsec:requirements} for any realistic imbalance of the surface photoemission properties.
	\item Minimize charge noise due to discharge actuation
\end{itemize}

\begin{figure}[!t]
\centering
\includegraphics[width=0.46\textwidth]{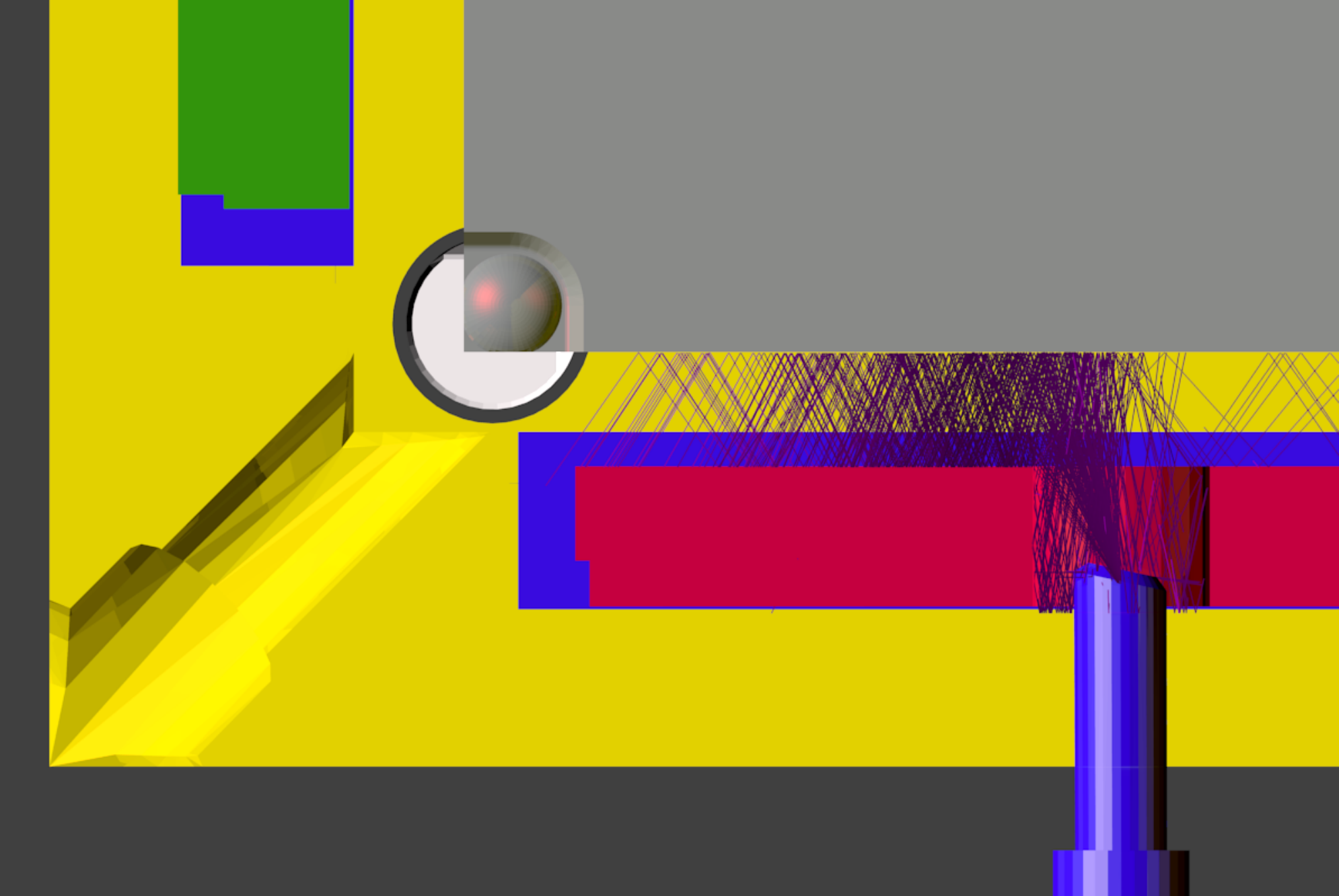}
\caption{Proposed light injection via an UVLF, located in the existing hole of the injection electrode (red), on the $-y$-side of the electrode housing (yellow). Some simulated UV light rays are shown. The UVLA tip with the fiber inside is chamfered by $30^\circ$, such that the central light beam is deflected by $18.8^\circ$.}
\label{fig:LISA_lightinjection}
\end{figure}

The upper goals can be achieved by two major design changes:
\begin{enumerate}
	\item \emph{Change of the UV light injection into the inertial sensor} such that the light is injected through the existing holes in the center of the injection electrodes which are mounted on the $y$-faces of the electrode housing (see Figure~\ref{fig:LISA_lightinjection}). The light shall be injected such that it is mainly restricted on the surfaces between the injection electrodes and its adjacent side on the test mass. 
	\item \emph{Change the UV light source and the actuation strategy} such that the light is only switched on when the present electric fields entirely suppresses the unwanted electron flow. 
\end{enumerate}

\begin{figure}[!t]
\centering
\includegraphics[width=0.46\textwidth]{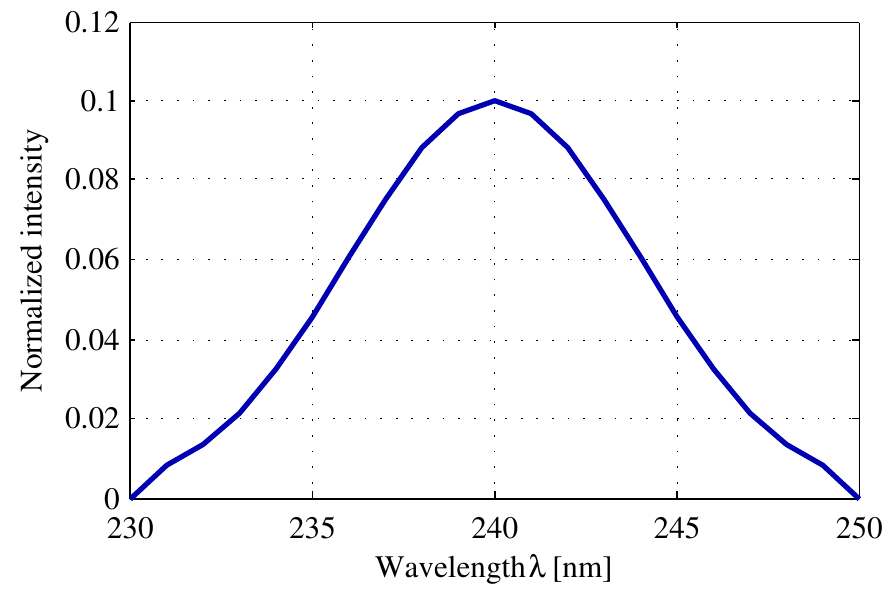}
\caption{Spectrum of the $240$\,nm UV LED as parameterized in the light source model of the ray tracer to generate random rays with different wavelengths.}
\label{fig:LED_spectrum}
\end{figure}

As indicated in Section~\ref{subsec:contamination}, higher UV light energies mitigate the effects of small work function changes due to molecular adsorption as well as effects related to surface emission. Thus, the light source characteristics of commercially available UV LEDs with a spectral peak wavelength of $240$\,nm ($5.17$\,eV), a maximum output power of $100$\,$\mathrm{\mu}$W, and a maximum modulation frequency of up to $100$\,MHz are assumed in the following calculations. The spectrum as parameterized in the light source model of the ray tracer is shown in Figure~\ref{fig:LED_spectrum}.
The effect of reduced wavelengths on photo-electron emission from air-contaminated gold surfaces is currently studied in the scope of an ESA technology development program \cite{esa_ITT_LISACMS}.\\

\subsubsection{Light injection}
\label{subsubsec:LISA_lightinjection}
It is proposed to inject the UV light through each of the two existing holes on the $+y$ and $-y$ side of the electrode housing (see Appendix~\ref{subsec:ISdescription}). As and example, Figure~\ref{fig:LISA_lightinjection} shows the UVLF located on the $-y$-side of the electrode housing.
The light injection has been optimized by running multiple ray-tracing simulations with the optimization criterion to maximize the illumination of the injection electrode on the $y$-face and its adjacent side on the test mass, while minimizing the illumination of other parts (e.g., the UVLF and the actuation/sensing electrodes). 

The core of the UVLF is a multimode silica fiber of $1$\,mm diameter with a refractive index $n=1.5$ at the peak wavelength (same fiber as used for LPF). The fiber tip is chamfered by $30^\circ$ such that the central beam of the exiting UV light is deflected away from the central axis of the UVLF by $18.8^\circ$. The deflection is caused by the change of refractive index at the transition between the silica fiber and vacuum, as described by Snell's law. In principle, this method allows deflection angles up to $\approx40^\circ$, unless the central ray of the light beam is totally reflected inside the fiber. However, for the expected beam spread, parts of the outer beam wings are already reflected when the chamfer angle exceeds $30^\circ$. Refractive light deflection is used, because it allows to sufficiently fulfill the optimization criterion and has a reduced manufacturing complexity compared to a micro-optical element.

From the visualization of the ray tracing simulation in Figure~\ref{fig:LISA_lightinjection}, it becomes clear that the light initially hits the test mass and is then reflected back to the injection electrode. Consequently, $71.5$\,\% of the light will be absorbed by the test mass and $23.2$\,\% by the adjacent injection electrode; the remaining $5.3$\,\% are absorbed by other surfaces (mainly by the UVLF and the hole in the injection electrode). Note that the light distribution within the sensor is constrained on regions with simple geometry and only one dominant electric field (caused by the voltage applied to the injection electrode and by the test mass potential).\\

\subsubsection{Actuation strategy}
\label{subsubsec:LISA_actuationstrategy}
\begin{figure}[!t]
\centering
\includegraphics[width=0.46\textwidth]{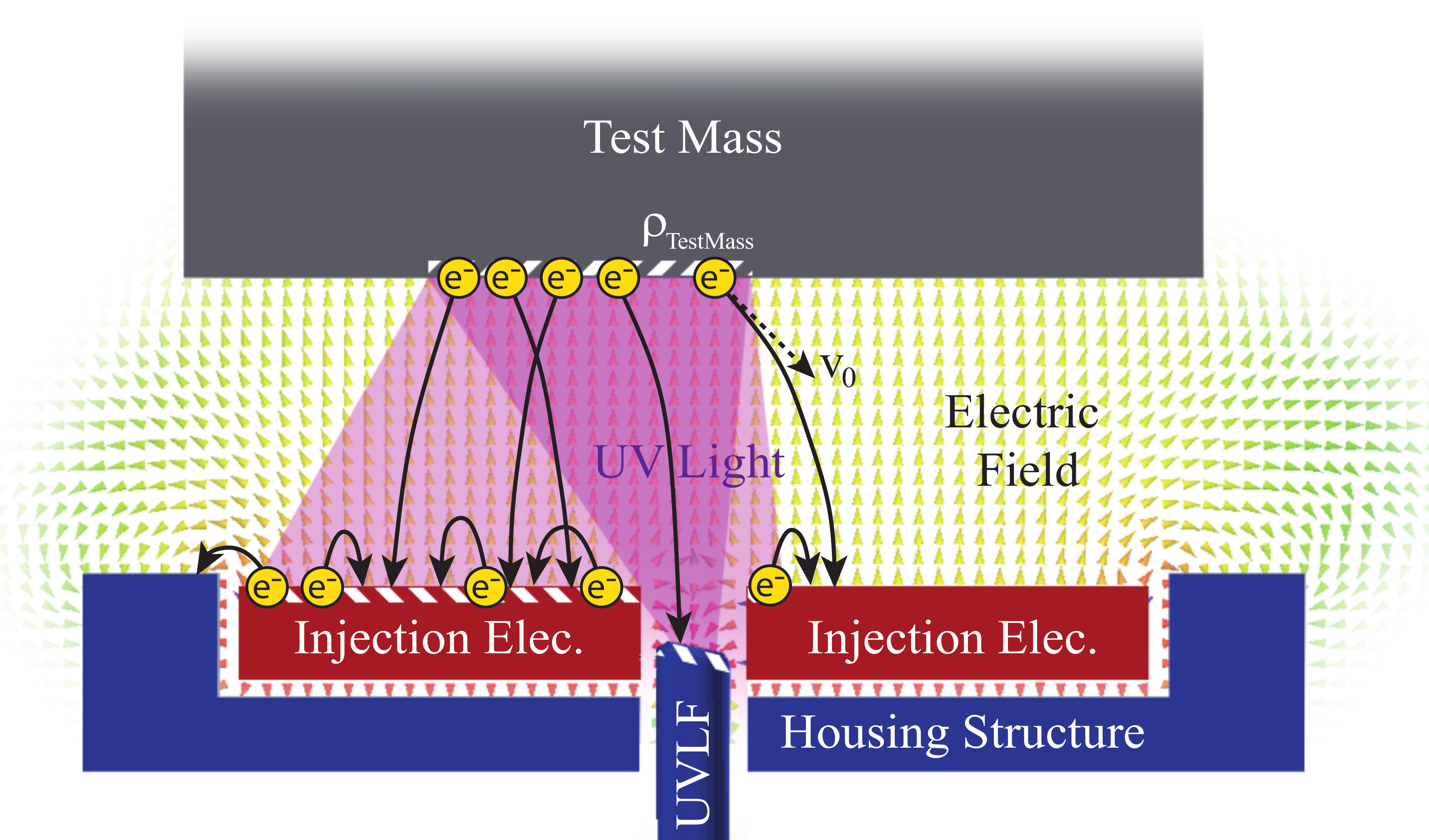}
\caption{Schematic illustration of the pulsed-light-source design, shown for a positive test mass discharge rate. The light pulses are synchronized with the injection voltage such that emitted electrons from the EH are suppressed by the electric field and those from the TM are accelerated towards the EH.}
\label{fig:LISA_discharge_principle}
\end{figure}

Due to the obtained light distribution, electrons are always emitted from the test mass and the electrode housing parts (similar to the unsynchronized DC charge control principle of LISA Pathfinder). Depending on the kinetic energy of the photo-electrons and the electric fields in the gap between injection electrode and test mass, they either reach the adjacent side or are attracted back to the emitting side. A robust \emph{positive} test mass discharge rate is obtained by switching on the light only when the electric field is such that all electrons emitted from the EH are suppressed (and the transition of electrons emitted from the test mass is enhanced), and, vice versa for negative discharge rates. Figure~\ref{fig:LISA_discharge_principle} illustrates the ``UV light pulsing'' principle for a positive test mass discharge rate. The schematic shows the UV light injection via the chamfered UVLF, the instantaneous electrical field situation between the illuminated test mass and electrode housing surfaces, and the desired electron trajectories to achieve positive discharge rates. 

\begin{figure}[!t]
\centering
\includegraphics[width=0.46\textwidth]{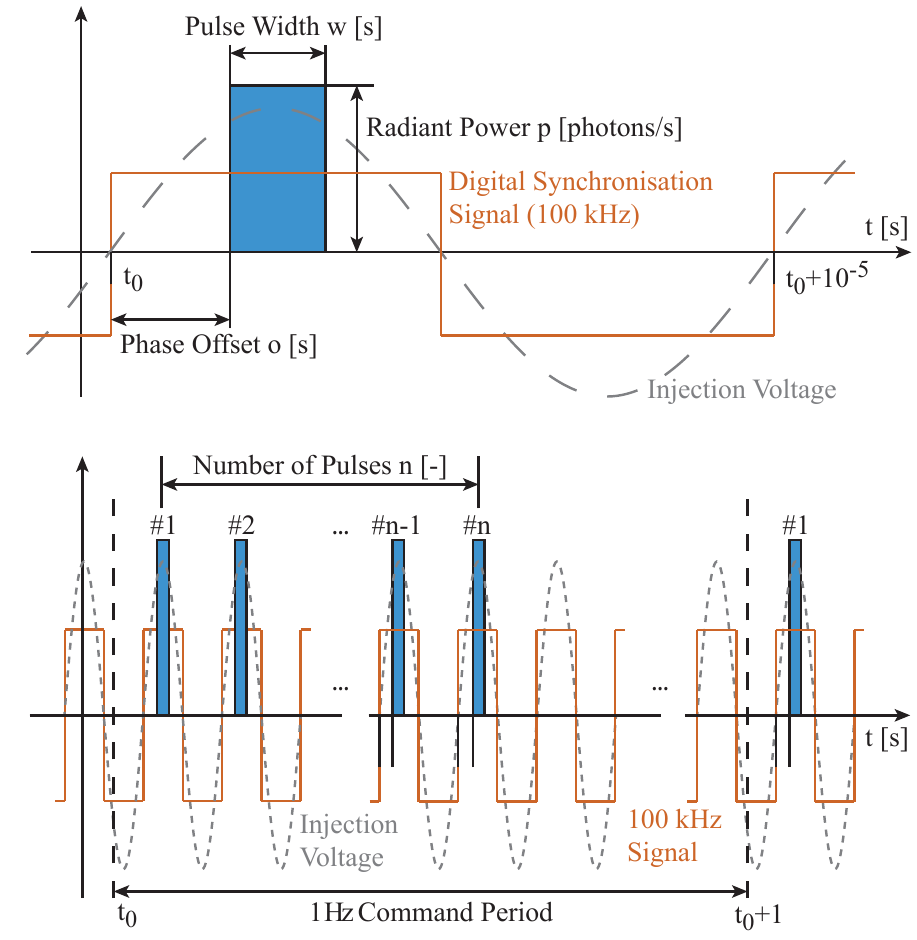}
\caption{Concept of proposed UV light switching (synchronized AC charge control). The four actuation commands ($n$, $p$, $o$, $w$) define the number of light pulses per $1$\,Hz command period, the radiant power of a single pulse, phase offset of the light pulses w.r.t. the $100$\,kHz synchronization signal, and the pulse width of the light pulses.}
\label{fig:ac_commands}
\end{figure}

Since no other electrodes than the $y$-injection electrodes are illuminated, the already present $100$\,kHz sinusoidal injection voltage (see Eq.~\ref{eq:injectionvoltage} in Appendix~\ref{subsec:ISdescription}) is adequate to properly control the transition of the emitted electrons from both sides. This requires the light pulses to be \emph{synchronized} with the $100$\,kHz injection voltage. The concept of the proposed UV light pulsing relative to a digital $100$\,kHz synchronization signal is shown in Figure~\ref{fig:ac_commands}. The digital synchronization signal is an input to the light source electronics and should be provided by the same actuation and sensing electronics which also generates the injection voltage.

Furthermore, in order to describe the synchronized on/off switching of the UV light in a compact way, a command set consisting of the four parameters ($n$, $p$, $o$, $w$) at a command rate of $1$\,Hz is proposed for the UV light source unit (see Figure~\ref{fig:ac_commands}). With this concept, the load of the spacecraft communication bus is minimized, while achieving a high dynamic range of the UV output. 

The parameter $n$ defines the number of light pulses per $1$\,Hz command period. The maximum number of light pulses is determined by the number of periods of the $100$\,kHz injection voltage within one second. Thus, between $0$ and $10^5$ light pulses can be generated for both the positive and the negative half-periods of the injection voltage. 
The parameter $p$ specifies the radiant power of the individual light pulses. 
The parameter $o$ sets the phase offset between the light pulses and the $100$\,kHz digital synchronization signal. Therefore, the offset parameter also defines whether a positive or a negative discharge rate is obtained (by ``shifting'' the light pulses to either the positive or the negative half-period of the injection voltage). Note that each of the two feedthroughs can be used to generate positive and negative discharge rates, providing double redundancy for each discharge direction. The parameter $w$ sets the pulse width of the individual light pulses. 

The resolution of $p$, $o$, and $w$ is proposed to be $8$\,bit, such that $256$ steps can be commanded for each of these parameters. Thus, the phase offset $o$ ranges between $0$ (i.e., no offset) and $255$ ($10\,\mathrm{\mu}$s offset), with a linear step size of $\approx40$\,ns. By specifying the maximum pulse width $w$ to be one half of the injection voltage period, the maximum pulse width command is $5\,\mathrm{\mu}$s, and the minimum possible pulse width command approximately $20$\,ns (which also defines the pulse width command resolution).

The resolution of the parameters $n$, $p$, and $w$ define the theoretical dynamic range of the UV light source according to $10^5\cdot2^8\cdot2^8\approx6.6\cdot10^9$. This is more than $7$ orders of magnitude larger than the dynamic range of the LISA Pathfinder UV light source.

In order to obtain bipolar discharge rates, independently from quantum yield imbalances between EH and TM surfaces, the offset and the pulse duration within one half-period of the injection voltage are selected such that its amplitude is always larger than the kinetic energy of the emitted electrons. Assuming the maximum electron kinetic energies to be $3$\,eV (which is a worst case assumption since it would correspond to work functions of only $2.1$\,eV), the phase offset and the pulse duration are set such that the light is only switched on when the voltage between the injection electrode and the TM is larger than $3$\,V. This reduces the theoretical dynamic range by approximately a factor of $2$, but causes the undesired electron flow to be almost entirely suppressed.

\subsection{Analysis of the Robust Discharge System}
\label{subsubsec:LISA_achieved_performance}
\begin{figure}[!t]
\centering
\includegraphics[width=0.46\textwidth]{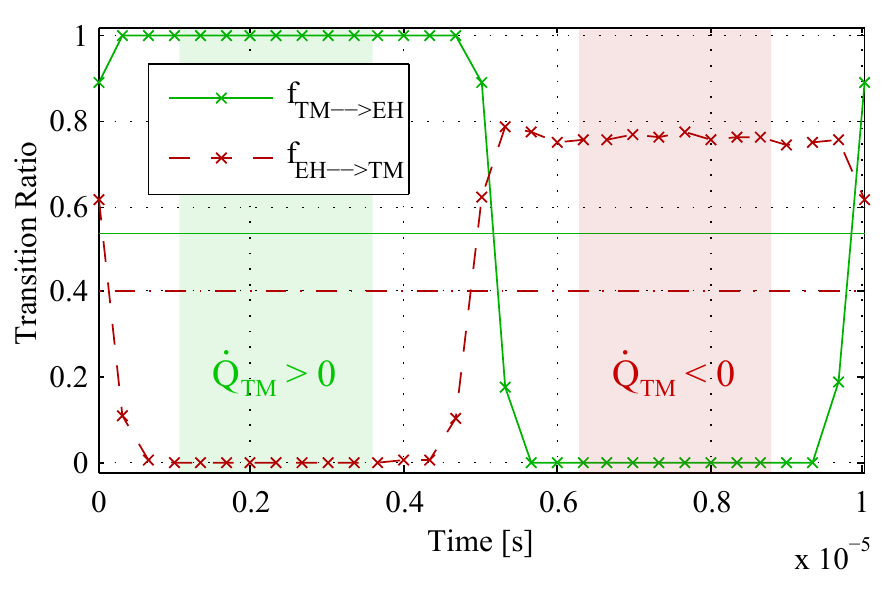}
\caption{Transition ratios of the emitted electrons between EH and TM surfaces over the period of the injection voltage. The ideal time slots in which the UV light has to be switched on to obtain positive and negative test mass discharge rates are highlighted.}
\label{fig:LISA_transition_ratios}
\end{figure}

Figure~\ref{fig:LISA_transition_ratios} shows the calculated transition ratios between the test mass and the electrode housing surfaces over the period of the $100$\,kHz injection voltage V$_\text{inj}$. The effect of the actuation voltages is considered since the shown transition ratios are averaged over all injection voltage periods within one second. 
The highlighted area in the center of the first half-period of the $100$\,kHz injection voltage shows that the transition of electrons from the EH to the TM is almost completely suppressed ($f_{\text{EH}\rightarrow \text{TM}}(t)\approx0$); the transition ratio from TM to EH is $f_{\text{TM}\rightarrow \text{EH}}(t)\approx1$. Thus, the UV light is switched on during this time slot to obtain positive discharge rates. 
On the other hand, the electrons emitted from the test mass are almost completely suppressed in the center of the second half-period ($f_{\text{TM}\rightarrow \text{EH}}(t)\approx0$); thus, the light is switched on during this time slot to obtain negative discharge rates. 

In both cases, the basic principle is to suppress the ``unwanted'' electron flow as much as possible. The quantum yield imbalances between EH and TM surfaces merely contribute to the \emph{uncertainty} in the amplitude of the obtained discharge rates, but do not influence the \emph{functionality} to obtain bipolar discharge rates. The desired discharge rate amplitudes can be recovered through the amount of injected UV light power (assuming the dynamic range of the light source is sufficiently large). 

Table~\ref{table_robustness_LISA} shows the robustness gains of the proposed design, without application of DC bias voltages. When the typical actuation voltages for zero force/torque command (see Appendix~\ref{subsec:ISdescription}) are considered in the analysis, bipolar discharge rates can be realized for quantum yield imbalances up to a factor of $\approx884$. The effect of the actuation voltages is indicated by the calculated robustness gains when \emph{no} actuation voltages are applied. 

\begin{table}[h!]
\caption{\label{table_robustness_LISA} Predicted robustness of the proposed discharge system design}
\begin{tabular}{@{}lcccc@{}} \toprule
\centering
           & \multicolumn{2}{c}{No Actuation Voltages} &  \multicolumn{2}{c}{With Actuation Voltages}   \\ \cmidrule{2-5}
DC Voltage Scenario  &  $\Delta \dot{Q}^+$ & $\Delta \dot{Q}^-$ & $\Delta \dot{Q}^+$ & $\Delta \dot{Q}^-$\\ \midrule
No DC volt.    									& 1238.6 & 785.2 &  1203.5  &  884.1 \\ \bottomrule
\end{tabular}
\end{table}

Moreover, electron kinetic energies of about $3$\,eV (corresponding to a work function of $2.14$\,eV) can be tolerated without degradation of discharge performance. This assumes that the phase offset $o$ and the pulse duration $w$ are such that the UV light is only switched on when the undesired electron flow is sufficiently suppressed (see Figure~\ref{fig:LISA_transition_ratios}).

Assuming a quantum yield imbalance of $884$ (corresponding to smallest acceptable value of the robustness gains in Table~\ref{table_robustness_LISA}, when actuation voltages are applied), the discharge performance requirements of Section~\ref{subsec:requirements} can be fulfilled when the dynamic range of the light source is I$_\text{UV,max}$/I$_\text{UV,min}=1.6\cdot10^7$. This range can be achieved with the presented design.

\section{Conclusion}
A comprehensive modeling toolbox for contact-free UV light discharge systems has been developed and discussed. The results from dedicated measurement campaigns have been used as model inputs (e.g., width and shape of the kinetic energy distribution of the photo-electrons, quantum yield, reflection curves) to have a representative description of the underlying physics. The toolbox has been used to analyze the existing flight hardware of the LISA Pathfinder discharge system. Important findings are that the robustness of the baseline design is marginal, since it does not guarantee the capability to safely produce bipolar discharge rates for the observed variations of the measured surface properties. These findings motivated the definition of stringent AIT requirements, the optimization of the UV light injection and of the test mass coating, with the goal to increase the system robustness. The optimization has been performed by means of extensive simulations using the developed discharge toolbox with inputs from dedicated measurement campaigns.

An analytical discharge model has been derived on the basis of the calculated surface illumination ratios and the electron trajectories. In addition to quantify the robustness of the LISA Pathfinder discharge system, such a model is used for the development of the charge control on-board software algorithms. Moreover, the model is the core of the discharge performance part of the overall LISA Pathfinder performance budget.

The discharge toolbox has been used to design and optimize a more robust UV light discharge system, which is directly applicable to cubical test mass inertial sensors as used in LPF and planned to be used in LISA/NGO. The analysis of the proposed design shows a significant increase of the robustness and performance figures. The concept is the baseline discharge system design for LISA/NGO as defined in the scope of the LISA Mission Formulation Study \cite{gath_LISA_formulation_study_overview2009}. Furthermore, from the new design and its detailed analysis, performance requirements for the needed UV light source have been derived. These requirements are inputs to a UV light source technology development program, currently running under ESA contract \cite{esa_ITT_LISACMS}.

\appendices
\section{Inertial Sensor Description}
\label{subsec:ISdescription}
In the following, the main components of space inertial sensors with cubical test masses are introduced, as far as they are needed for the understanding and modeling of UV light discharge systems. 

\begin{figure}[!t]
\centering
\includegraphics[width=0.46\textwidth]{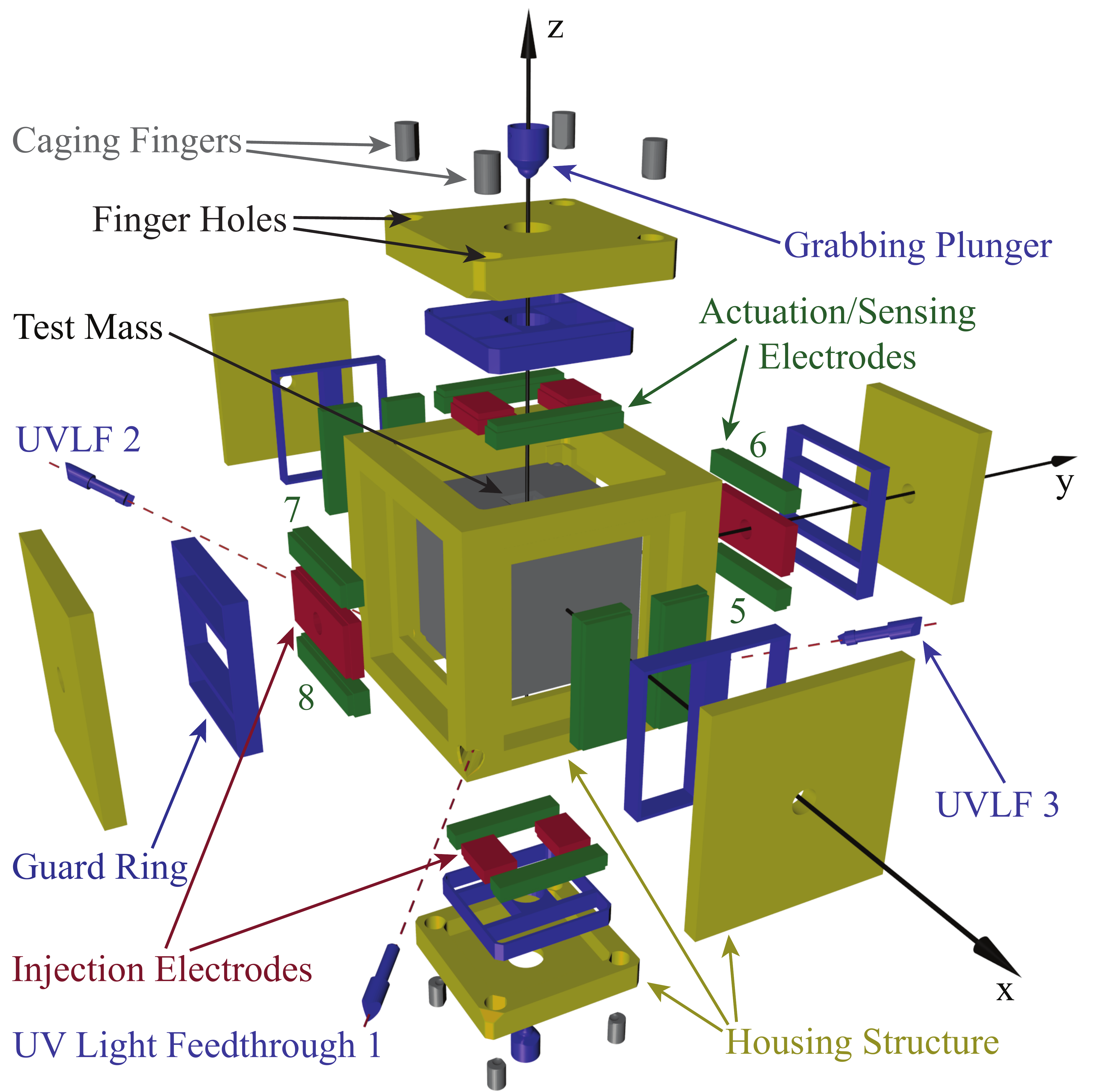}
\caption{Exploded view of the LISA Pathfinder inertial sensor including the cubical test mass, the electrode housing structure with the actuation/sensing, and injection electrodes, the caging fingers from the launch lock device, the grabbing plungers from the release mechanism, and the injection feedtroughs for the UV light optical fibers.}
\label{fig:LPF_IS_Explosion}
\end{figure}

Figure~\ref{fig:LPF_IS_Explosion} shows an exploded view of the inertial sensor which has been used as the reference scenario for the presented article. Two of these sensors are part of the scientific payload (LTP) on board the LISA Pathfinder spacecraft. The cubical, $1.96$\,kg and $46\times46$\,mm sized gold-platinum (Au$_{73}$Pt$_{27}$) test mass is nominally centered in the housing structure that carries the actuation/sensing and injection electrodes. All electrodes are surrounded by grounded guard rings to reduce fringe field effects. The inner parts of the sensor (except the spherical interfaces at the test mass corners and the caging fingers) are coated with a gold layer of $800$\,nm thickness on top of a $200$\,nm thick titanium layer. Because of the large gaps between test mass and electrodes ($4$\,mm along the sensitive $x$-axis), the test masses have to be caged during launch and will be released into free flight as soon as the spacecraft has reached the operational orbit. The release procedure is performed in two steps: First, the 8 iridium caging fingers (which apply the high loads during launch) are retracted along the $z$-axis into the finger holes, such that the test mass is held only by the two grabbing plungers. In a second step, the plungers are then simultaneously retracted such that the test mass is separated from their release tips\footnotemark\footnotetext{When two metallic bodies (like the gold coated test mass and gold-platinum dental alloy of the release tip) have been brought into contact and are then separated, two opposite charges remain on the two bodies \cite{contact_electrification_metals} such that a random charge Q$_\text{TM}$ with arbitrary sign will be left on the test masses.}.

On each side of the electrode housing (EH), two actuation/sensing electrodes are mounted. These electrodes are used to apply electrostatic forces and torques along each test mass degree of freedom (DoF) and to measure the linear and angular test mass displacement via the change of test mass to electrode capacitance. 

For the actuation of individual test mass DoFs, always a set of four electrodes is used (electrodes 1--4 for actuation of the test mass coordinates $x$ and $\phi$, 5--8 for $y$ and $\theta$, and 9--12 for $z$ and $\eta$). The actuation voltages are sinusoids with different frequencies for different test mass DoFs. 
Eq.~\ref{eqn:suspension_voltages} shows the actuation voltages on the $y$/$\theta$-electrodes as an example:
\begin{eqnarray}
 \label{eqn:suspension_voltages} 
V_{\text{act},5} &=& +A_{1y}(t)\sin(2\pi f_{y}t) + A_{1\theta}(t)\sin(2\pi f_{\theta}t)  \nonumber \\
V_{\text{act},6} &=& -A_{1y}(t)\sin(2\pi f_{y}t) + A_{2\theta}(t)\cos(2\pi f_{\theta}t) \nonumber \\
V_{\text{act},7} &=& +A_{2y}(t)\cos(2\pi f_{y}t) - A_{1\theta}(t)\sin(2\pi f_{\theta}t) \nonumber \\
V_{\text{act},8} &=& -A_{2y}(t)\cos(2\pi f_{y}t) - A_{2\theta}(t)\cos(2\pi f_{\theta}t)
\end{eqnarray}
The voltage amplitudes $A_{iy}$ and $A_{i\theta}$ are computed from the forces and torques as commanded by the DFACS on-board control software and depend on the disturbances occuring during spacecraft operation. The frequencies of the shown actuation voltages are $f_y=90$\,Hz and $f_\theta=180$\,Hz. The frequencies of all six actuation voltages range between $60-270$\,Hz and are orthogonal within one DFACS actuation cycle ($10$\,Hz), to minimize electrostatic force and torque cross-couplings.

Also sinusoidal test signal voltages $V_{\text{TS},i}$ for system identification purposes (e.g., test mass charge estimation \cite{ziegler_PrinciplesLPFDischargeSystem}) and DC bias voltages $V_{\text{DC},i}$ (e.g., to assist test mass discharging) can be applied via the on-board software to each of the twelve actuation/sensing electrodes. The maximum amplitudes of the DC bias voltages are $\pm5$\,V. The frequency is restricted by the $10$\,Hz sampling frequency of the on-board computer. In total, the following voltages can be applied to each actuation/sensing electrode: 
 \begin{eqnarray}
\label{eq:actsens_voltage}
V_{\text{EH},i} =  V_{\text{act},i} + V_{\text{TS},i} + V_{\text{DC},i} & i=1\ldots12.
\end{eqnarray}

The six injection electrodes (located between the actuation and sensing electrodes on the $y$ and $z$ faces of the electrode housing) are used to apply high frequency sinusoidal voltages to bias the test mass for electrostatic sensing purposes. The injection voltage is equal for all six injection electrodes:
\begin{eqnarray}
\label{eq:injectionvoltage}
V_{\text{inj},j} =  4.82 \cdot \sin(2\pi f_\text{inj}\cdot t) & j=1\ldots6,
\end{eqnarray}
where $f_\text{inj}=100$\,kHz. The potential of the free-floating test mass is given as:
\begin{equation}
\label{eq:TMpotential}
V_\text{TM}=\frac{1}{C_\text{tot}} \Big( Q_\text{TM}+\sum_i C_{\text{EH},i} V_{\text{EH},i}+\sum_j C_{\text{inj},j} V_{\text{inj},j} \Big),
\end{equation}
where $C_{\text{EH},i}$ and $C_{\text{inj},j}$ are the individual electrode capacitances, $C_\text{tot}$ the total test mass to housing capacitance, and $Q_\text{TM}$ the accumulated test mass charge.

The electric field, caused by the electrode voltages and by the test mass potential, affects the trajectories of the photo-electrons, emitted from electrode housing and test mass surfaces under UV illumination. The UV light is injected into the sensor through dedicated holes at the corners of the EH $-z$ side by means of UV light feedthroughs (UVLF) with optical fibers inside. There are $3$ UVLFs, two directed towards the electrode housing and one towards the test mass. The fibers are connected to a UV light source (fibers and light source are not shown in Figure~\ref{fig:LPF_IS_Explosion}).

\section*{Acknowledgment}
The authors thank Andr\'{e} Posch, Federico Pinchetti (both University of Stuttgart), and Duy C. Nguyen (University of Applied Sciences Darmstadt) for their invaluable support in the development of the ray tracing and electron tracing tools. We further thank Markus Pfeil (former Imperial College London, now TWT GmbH) for his support in the early days of our involvement in the topic, Ulrich Johann and R\"udiger Gerndt (both Astrium GmbH) for their advice in many important discussions. The work was partially performed under contracts from ESA and DLR. We thank C\'{e}sar Garc\'{i}a Marirrodriga, Giuseppe Racca, Bengt Johlander, Alberto Gianolio (all ESA), and Hans-Georg Grothues (DLR) for their support and encouragement in these activities.

\ifCLASSOPTIONcaptionsoff
  \newpage
\fi

\bibliographystyle{IEEEtran}
\bibliography{biblio_discharge}

\end{document}